\documentclass[fleqn,10pt]{olplainarticle}

\usepackage{longtable}
\usepackage{array}

\title{Synthetic Human Mobility Data Generation: A Structured Review of Representations, Methods, and Practical Capabilities}


\begin{abstract}
Human mobility data has become an increasingly important component of urban analytics. Although the range of available mobility data sources has expanded substantially, access remains highly constrained by commercial restrictions, privacy concerns, and institutional barriers. Data protection procedures also often reduce the analytical value of released datasets.

Synthetic mobility data has emerged as a promising solution, but existing methods differ substantially in their underlying mechanisms, the information they preserve, the outputs they generate, and the analytical questions they can support. Their comparative strengths and trade-offs remain insufficiently understood for urban analytics applications.

This paper presents a structured review of synthetic human mobility data generation from an urban analytics perspective. We review the literature by methodological family, and index it by the mobility outputs each family generates natively and the analytical capabilities those outputs enable. We first provide a taxonomy of synthetic data products---including population and persona representations, activity schedules, trip and tour records, trajectories, and aggregate mobility patterns. We then review the major methodological families, spanning mechanistic models, survey-driven population synthesis, activity- and agent-based simulation, deep generative models, transformer-based mobility language models, and LLM-agentic systems. Building on this synthesis, we introduce a Meaning--Population--Autonomy framework that characterises these methods along three dimensions: behavioural meaning, population grounding and scale, and generation autonomy. We consider these dimensions the principal requirements for downstream urban analytics. Few methods deliver behavioural meaning, population grounding and autonomous generation at once, and fewer still with generation constrained to feasible trajectories.

\end{abstract}

\author[1]{Yanbo Pang}
\author[1]{Chen Zhong}
\author[2]{Song Gao}
\author[3]{Yoshihide Sekimoto}
\affil[1]{Centre for Advanced Spatial Analysis, University College London}
\affil[2]{Geospatial Data Science Lab, Department of Geography, University of Wisconsin-Madison}
\affil[3]{Center for Spatial Information Science, the University of Tokyo}

\begin{document}

\flushbottom
\maketitle
\thispagestyle{empty}

\section{Introduction}
\label{sec:intro}

Human mobility is one of the most direct ways to observe how cities function in practice. Human mobility data does not merely record movement: it provides partial but valuable evidence on how people traverse urban space and time and, when combined with contextual information, can help infer how different parts of the city are connected through daily activity \citep{kwan2012uncertain,Zhao2016,Barbosa2018,Moro2021,xu2025predicting}. This makes mobility central to urban analytics, as it redefines how cities are understood by revealing their functional organisation through recurring patterns of travel, activities, interactions, and constraints. Those patterns are strongly regular: travel displacements follow heavy-tailed distributions, individuals return repeatedly to a small number of significant places, and daily movement is substantially predictable despite its apparent randomness \citep{Brockmann2006,Gonzalez2008,Song2010}. These empirical regularities matter directly for synthetic mobility, because they are simultaneously what generative models must reproduce and the yardstick against which generated data is judged realistic. As a result, human mobility data has become a cornerstone for analysing accessibility, spatial inequality, commuting structure, functional urban areas, temporal demand patterns, infrastructure pressure, epidemic spread, disaster response, and the behavioural consequences of planning or policy interventions \citep{toole2012inferring,Tizzoni2014,Xu2018,Haraguchi2022}.

Human mobility data gained unprecedented prominence, particularly during the COVID-19 pandemic, for monitoring and modelling epidemic transmission \citep{gao2020association,Haraguchi2022}. Although the value of human mobility data is now widely recognised and such data is increasingly collected, significant barriers remain: such data is often difficult to access, share, and interpret. First, real mobility data is fragmented across sources that capture different aspects of movement, making them difficult to interpret as complete records of urban behaviour. Second, many of these sources are difficult to access and share because of privacy risks, commercial ownership, legal restrictions, and institutional agreements.

Contemporary human mobility datasets may be collected from smartphone GPS traces, mobile phone call detail records or signalling data, smart-card records, travel surveys, location-based services, social media, or sensor-based traffic monitoring. Each source provides a partial view. Travel surveys are behaviourally rich and record trip purpose, mode, traveller attributes, and household context, but are typically limited in sample size, update frequency, and spatial and temporal detail \citep{StopherGreaves2007}. Smartphone GPS traces and other passively collected records provide detailed spatiotemporal observations, yet often contain limited information about activity purpose, traveller attributes, or trip semantics \citep{ShenStopher2014}. Mobile phone data can offer broader population coverage, but usually at coarser spatial or temporal resolution and with limited socioeconomic status and behavioural annotation \citep{Blondel2015,Xu2018}. Smart-card data record only part of the travel process, often with strong coverage for specific public transport systems but weak visibility beyond them \citep{Pelletier2011}.  Location-based services and social media add further place-based and behavioural signals, but are shaped by platform-specific bias, uneven user coverage, and interpretive uncertainty \citep{Zheng2011,Marti2019}. These representational limitations are compounded by access and governance constraints, especially for fine-grained individual mobility records \citep{JiangPrivacy2022}. As a result, while real mobility data is indispensable for observing cities in motion, it is often difficult to assemble into openly shareable, reusable, and comparable research infrastructure.

These limitations have led to growing interest in \textbf{synthetic data: artificially generated data designed to reproduce selected statistical, structural, or behavioural properties of real-world data while reducing the need to expose the original individuals or records} \citep{rubin1993statistical,reiter2005releasing, Jordon2022,EuropeanDataProtectionSupervisor2025}. Synthetic human mobility data is valuable not only as a privacy-preserving substitute for sensitive location records, but also as a research infrastructure that supports downstream mobility analysis and urban applications. By generating artificial populations, activity schedules, trajectories, or aggregate flows under transparent modelling assumptions, synthetic mobility data can support reproducible method development, task-specific benchmark design, and controlled scenario analysis for urban and transport applications, while reducing direct dependence on proprietary or individually identifiable records~\citep{JiangTimeGeo2016,yabe2024enhancing,Kapp2024}. Yet synthetic human mobility is not a simple extension of synthetic tabular, image, or text generation. Mobility is multi-layered: socioeconomic, demographic, or persona-based conditioning, daily activity schedules, trip-level representations, trajectories, and aggregate mobility outputs are analytically distinct, yet closely linked in practice \citep{muller2010population,Felbermair2020,Kashiyama2024, Kapp2024}. A valid synthetic mobility dataset must keep several layers and attributes aligned. Individual patterns should look believable in how people actually move; trips should be possible given the city's layout, transport networks, points of interest (POI), and timing constraints~\citep{Zhu2024}; and the full dataset should recover broader urban regularities such as density patterns, origin--destination (OD) structure, population distributions, and daily rhythms \citep{Kong2023,Rao2023,Kapp2024}. Recent research uses satellite imagery that can capture urban contexts to generate population-level human mobility flows in cities~\citep{rong2026satellites,wang2026sat2flow} and fuse remote sensing and social sensing data to bridge the representation gap of single data modality~\citep{xu2026multimodal}.

For many urban and social researchers, transport planners, and policy analysts, the starting point is an analytical need rather than an algorithmic preference. A study may require daily activity schedules for behavioural analysis, trip and tour records for travel-demand analysis, trajectories for route or exposure analysis, synthetic people-flow for population-level applications, or aggregate mobility outputs for infrastructure planning. Users also need to know whether a method represents a target population, whether it requires observed individual histories, and whether the required output is generated natively or obtained only after grounding, simulation, refinement, or aggregation.

A further challenge is that human mobility is not only about where and when people move, but also about why they travel, what activities those movements represent, and what is the rationale for their choices and what are their constraints, as outlined in the time geography framework~\citep{miller2017time,neutens2011prism,shaw2023time}. Researchers and practitioners seek to understand the drivers and mechanisms underlying human mobility, rather than viewing it merely as sequences of location points~\citep{McFadden1974,BenAkivaLerman1985, AxhausenGarling1992,BowmanBenAkiva2001,Nayak2023}, as well as to integrate multi-source and multi-modal data.

Existing methods therefore differ in what they treat as the primary object of generation. Survey-driven, diary-based, activity-based, and agent-based approaches often produce behaviourally structured outputs, such as daily activity schedules, or demographic attributes, before deriving trips or trajectories \citep{Pappalardo2018,Felbermair2020,Nayak2023}. Recent large language model (LLM)-based and agentic workflows further extend this semantic direction by generating travel diaries, plans, or persona-conditioned mobility structures \citep{WangUrbanResidents2024,LiMobAgent2024,Liu2026}. By contrast, many trajectory-oriented deep generative models that emerged in the GeoAI era, including generative adversarial network (GAN)-based, diffusion-based, and transformer-based variants, primarily generate spatiotemporal traces or mobility tokens, even when semantic signals are used as inputs or controls \citep{Jiang2023,Rao2023,ZhuDiffTraj2023,chu2024simulating,Song2024,yuan2026worldmove}. Methods also differ in whether they generate individual traces, dataset-level synthetic outputs, or population-grounded mobility, and whether they rely on synthetic populations, personas, aggregate priors, partial sequences, or observed user histories at generation time \citep{Berke2022,Felbermair2020,Kashiyama2024,Hsu2024}.

Most reviews on synthetic data generation focus on the development of computational methods and highlight algorithmic preferences \citep{Barbosa2018,Kong2023,Kapp2024,Luca2021}. Although useful, their guidance for downstream urban applications is limited. End users, such as urban and social researchers, transport planners, and policy analysts, often consider data from different perspectives, e.g., how useful it is in the context of the applications, rather than how technically advanced an algorithm is. For instance, they may prioritise trajectories for exposure or route analysis, daily activity schedules for behavioural interpretation, trip purposes and transport modes for modal-shift scenarios, population-grounded synthetic mobility for distributional analysis, or aggregated people and traffic flows for infrastructure planning. That said, a technically sophisticated method may still be unsuitable if it generates the wrong output form, depends on unavailable user histories, lacks population grounding, or cannot be evaluated at the scale required by the application.

This review addresses this gap. Rather than comparing computational model families on their own terms, we index them by the outputs and capabilities that determine practical use in urban and social sciences. We first \textbf{define} the main mobility representations that are most commonly used in human mobility-related urban analytics, then \textbf{review} the methodological families that generate them, emphasising the underlying mechanism in human mobility data generation as it matters to what feature has been preserved in synthetic data, and finally \textbf{compare} methods through three capability requirements: human behavioural meaning, population grounding and scale, and generation autonomy, which we consider to be the principal elements for supporting analytical tasks. The goal is to help downstream data users identify which methods are suitable for which urban analytics needs, and where current methods still fail to provide practical synthetic mobility infrastructure. Specifically, we organise the review around the following practical questions.

\begin{itemize}
    \item What types of synthetic human mobility outputs are most commonly used for urban analytics tasks, and how can they be summarised using a unified representational terminology? 
    \item Which methods can produce each output form, and what do they generate natively?
    \item What practical capabilities do these methods offer, particularly with respect to human behavioural meaning, population grounding and scale, and generation autonomy?
        
\end{itemize}

These questions define the use-oriented structure of the review. Alongside model family, we use output form, native generation logic, and practical capability as indexing dimensions. This allows the review to serve not only as a summary of existing methods, but also as a guide for readers seeking synthetic mobility data or methods suited to particular urban-analytics tasks.

The remainder of this paper is organised as follows. Section~\ref{sec:scope} defines the scope of the review and clarifies the representational terminology used throughout the paper. Section~\ref{sec:litcollection} outlines the literature collection and review strategy. Section~\ref{sec:families} reviews the major methodological families of synthetic human mobility generation. Section~\ref{sec:mpa} introduces the Meaning--Population--Autonomy requirements for practical large-scale synthetic mobility. Section~\ref{sec:challenges} identifies open challenges and research gaps, and Section~\ref{sec:conclusion} concludes.

\section{Scope and Terminology}
\label{sec:scope}

\subsection{Scope and task boundaries}

This review focuses on methods that generate artificial records intended to represent human mobility or mobility-related population dynamics. We include work when the generated output represents mobility at an individual, population, or aggregate level, rather than only when it produces GPS-like trajectories. The review is therefore narrower than general synthetic data generation, but broader than trajectory generation alone \citep{Jordon2022,Kapp2024,Kong2023}. Spatial scale bounds the review as well. We cover human mobility across a city and its surrounding functional region, the scale at which population coverage and urban spatial structure shape the generation problem, and it is in this sense that the review is directed at urban analytics. Movement confined to a single site, such as a building or a campus, or to a single neighbourhood, falls outside this scope. The main output forms covered by this scope are defined in the next subsection.

Synthetic mobility generation should also be distinguished from privacy protection. Synthetic data may reduce data-sharing risk by avoiding direct release of original individual records, but it does not automatically guarantee privacy \citep{stadler2022synthetic}. Privacy-preserving trajectory publishing, anonymisation, differential privacy, and location-privacy mechanisms are therefore not treated as a main organising axis of this review.  However, they are discussed when relevant, particularly when they directly produce synthetic mobility records, when they motivate a generation method, or when privacy is relevant to evaluating the risks and utility of generated data \citep{Jordon2022,Kapp2024,JiangPrivacy2022}.

Several other related prediction and optimisation tasks are also excluded from the main scope, because their objective is to predict outcomes or optimise existing movement rather than to regenerate data. Section~\ref{sec:collection} lists these tasks and states how each was handled during corpus construction. Nevertheless, some of them are touched upon as their modelling logic is explicitly adapted for synthetic mobility generation, infilling, or construction of artificial mobility records. Adjacent literatures that help define aggregate mobility outputs are treated the same way.

This scope means that the review is not organised around one data type, privacy mechanism, or model architecture. Instead, we seek a unified data representation that maximises the utility of synthetic mobility data across downstream applications, provides a systematic understanding of the mechanisms underlying different methodological families, and ultimately enables users to select methods that are most appropriate for their intended applications.

\subsection{Core terminology and concepts}

Across the studies reviewed in this paper, human mobility is a broad and multi-faceted concept that has been defined and described by several recurring terminologies, such as visitation, flows, trajectories, and more recently, personas. These forms differ in spatial resolution, temporal detail, behavioural meaning, and level of aggregation. We use them as a common vocabulary for locating and comparing methods and their outputs. This vocabulary is grounded in travel behaviour theory, where travel is treated as a derived demand; in empirical data structures such as household travel surveys, GPS traces, smart-card records, and OD statistics; and in synthetic mobility studies that target different native outputs \citep{McNally2000,OrtuzarWillumsen2011,Nayak2023,StopherGreaves2007,ShenStopher2014,Pelletier2011,Pappalardo2018,Felbermair2020,Kashiyama2024,Kapp2024}.

\paragraph{Population-related representations.}
Population appears in three distinct ways in this review. \textit{Static synthetic population} refers to disaggregated synthetic individuals and households, each assigned sociodemographic attributes such as age, occupation, income and associated with locations they based at or visited. In transport applications, this tradition was developed to create the household and person records required by travel-demand microsimulation \citep{Beckman1996,GuoBhat2007,muller2010population}. These synthetic persons and households then serve as the decision makers in activity-based travel demand models \citep{BhatKoppelman1999,BowmanBenAkiva2001}, and more broadly as agent populations in social or land-use simulation settings \citep{Waddell2002,Chapuis2022}. \textit{Dynamic or de facto population distribution} refers to time-varying population presence across urban space, such as daytime population, ambient population, hourly population density, or grid-level population movement \citep{Deville2014,Martin2015,Bergroth2022}. \textit{Synthetic people-flow or mobility population} connects population priors to generated activities, trips, trajectories, or people-flow patterns \citep{Population247NRT2021,MurataSyntheticPopulation}. In short, it describes how population structure is translated into mobility.

\paragraph{Personas and behavioural profiles.}
Personas are interpretable behavioural profiles that complement rather than replace synthetic populations. In particular, they add patterns of behaviour by combining attributes such as demographics, preferences, constraints, routines, household roles, mobility styles, or barriers to using particular transport options. Transport planning has gradually adopted  persona-like segmentation to summarise heterogeneous travel needs, as shown by the UK Department for Transport's transport user personas and Transport for London's Transport Classification of Londoners \citep{DfT2023TransportPersonas,DfT2023PersonasTechnical,TfL2017TCoL}. In synthetic mobility generation, personas are especially relevant when used for the recent LLM-based and other agentic methods, in which personas are encoded as prompts or agent profiles for travel diary generation, personal mobility simulation, or activity planning \citep{WangUrbanResidents2024,LiMobAgent2024,Liu2026}. Persona-based conditioning can improve interpretability and control, and should be validated against travel surveys, segmentation evidence, and observed mobility patterns \citep{LutzPersona2025}.

\paragraph{Daily activity schedules.}
Daily activity schedules refer to the activity-level organisation of a person's day before movement is expressed as trips or trajectories. In activity-based travel demand modelling, people travel because they need to participate in activities distributed across space and time. A daily activity schedule typically includes activity type or purpose, timing, duration, ordering, location or zone, and the coupling between activities and trips; depending on the model, it may also include household role, joint activity participation, constraints, mode-related information, or recurrent routines \citep{AxhausenGarling1992,BhatKoppelman1999,BowmanBenAkiva2001}. This concept matters because it defines mobility as behaviour rather than movement alone. Activity-based, diary-based, and LLM-agentic methods may represent this level as activity plans, mobility diaries, semantic state sequences, tokenised daily plans, travel itineraries, or persona-conditioned schedules \citep{JiangTimeGeo2016,Pappalardo2018,Felbermair2020,LiMobAgent2024,WangUrbanResidents2024,Liu2026}.

\paragraph{Trip and tour records.}
Trip-level representations describe mobility as discrete trips rather than as continuous paths. Following travel-survey and transport-demand usage, a trip is treated here as a one-way movement from an origin to a destination, usually with a departure or arrival time, and often with trip purpose and transport mode \citep{StopherGreaves2007,McNally2000,OrtuzarWillumsen2011}. A tour or trip chain organises multiple trips into an ordered daily travel structure, often around home or another anchor location \citep{McNally2000,OrtuzarWillumsen2011,BowmanBenAkiva2001,Ballis2020}. \textit{Travel mode} is treated as a trip-level behavioural attribute because it is attached to a travel episode and depends on OD structure, accessibility, car ownership, service supply, timing constraints, and person-level attributes \citep{McFadden1974,BenAkivaLerman1985,HensherRoseGreene2015}.

\paragraph{Trajectories.}
Trajectories represent movement at the path or trace level. At this level, mobility is described not only by origins, destinations, and travel times, but by the spatiotemporal path through which movement is realised. A trajectory may correspond to a single trip, a tour, or a longer daily mobility record, depending on the data source and modelling setting. In GIScience and GPS-based mobility research, trajectories are commonly represented as ordered spatial positions with timestamps, such as GPS point sequences, map-matched paths, road-link sequences, or other network-grounded traces \citep{ShenStopher2014,Kong2023}. In sequence-modelling work, the same movement may instead be discretised into tokens, defined below. Compared with trip and tour records, trajectories provide greater spatial detail about route choice, speed, exposure, local congestion, and spatial feasibility. At the same time, trajectories often contain weak behavioural meaning unless activity purpose, transport mode, or traveller attributes are inferred or attached through additional processing.

\paragraph{Aggregate mobility outputs.}
Aggregate mobility outputs describe mobility at the population or system level. They are the form in which mobility most often reaches official statistics and planning practice, as published tabulations rather than as the microdata behind them. They normally include OD matrices, inflow and outflow, link traffic counts, POI visitation counts, public-transit ridership, station entries and exits, temporal demand series, and population distributions, among others. These outputs are central to transport planning and urban analytics \citep{Burgalat2026} and can be directly linked to real-world planning questions, for instance, how many people use a corridor or station, and how population presence changes across the day \citep{Wilson1970,McNally2000,OrtuzarWillumsen2011,Pelletier2011,Deville2014,Martin2015,Bergroth2022}. Aggregate outputs may be directly modelled, as in spatial-interaction, trip-distribution, dynamic population mapping, or crowd-flow models, or derived by aggregating synthetic individuals, trips, or trajectories \citep{ZhangSTResNet2017,Pappalardo2018,Felbermair2020,Kashiyama2024}. They therefore serve both as native outputs in some methods and as macro-level validation targets for finer-grained generators.

\smallskip

\paragraph{Process and output terminology.}
Across these output forms, we use several terminologies consistently. In sequence-modelling contexts, a \textit{mobility token} refers to a discrete symbolic unit used to encode an element of mobility, such as a grid cell, road link, POI, time interval, activity type, or transport-mode state. \textit{Behavioural meaning} refers to interpretable information that explains mobility as purposeful action rather than merely as statistical movement patterns. \textit{Native output} refers to what a method directly produces; \textit{conditioning information} refers to contextual or constraint information supplied to guide generation; and \textit{derived outputs} refer to results obtained after additional processing. We reserve process terms for operations that connect outputs: \textit{grounding} links abstract plans or generated sequences to concrete places, networks, or schedules; \textit{refinement} repairs or adjusts generated traces; and \textit{aggregation} summarises individual mobility into population-level patterns; \textit{tokenisation} refers to converting continuous or structured mobility records into sequences of such units \citep{Pappalardo2018,Kashiyama2024,WangUrbanResidents2024,LiMobAgent2024,ShenStopher2014,Si2024,Wan2025}.

\begin{figure}[ht]
\centering
\includegraphics[width=\linewidth]{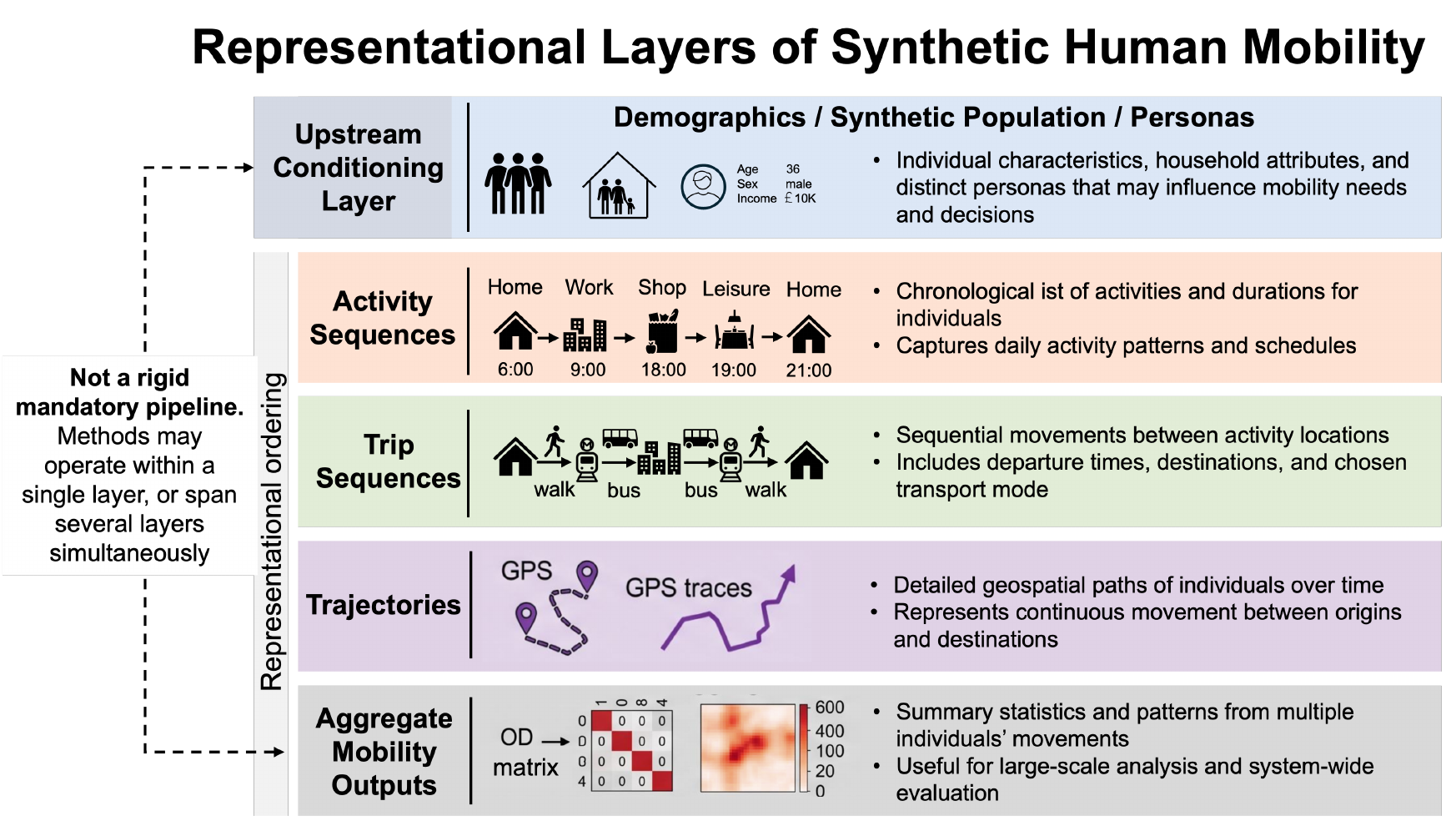}
\caption{Representational layers of synthetic human mobility. The figure organises synthetic human mobility into one upstream conditioning layer---demographics, synthetic populations, and personas---and four downstream representational layers: activity sequences, trip sequences, trajectories, and aggregate mobility outputs. This ordering reflects the behavioural logic of mobility: demographic and household attributes shape behavioural heterogeneity; activities create the need to travel; trips connect activities across locations; trajectories realise those trips in space and time; and aggregate patterns emerge only after many individuals are combined. The figure should not be interpreted as a rigid mandatory pipeline, since many methods operate at only one layer or link only a subset of layers. Rather, it provides a common representational framework for comparing how different methods describe, generate, or connect mobility across levels. ChatGPT and PicDoc were used to generate icon elements for this figure and to refine its layout; the authors reviewed the content and take full responsibility for it.}
\label{fig:mobility_representations}
\end{figure}

\paragraph{Relations among mobility representations.}
Taken together, these recurring forms can be organised through a common behavioural and analytical interpretation of mobility. Synthetic populations and behavioural profiles specify who is represented; daily activity schedules organise what people do; trip and tour records connect activities across locations; trajectories describe how those trips are realised in space and time; and aggregate mobility outputs describe population-level patterns, either through direct modelling or by combining finer-grained records. This interpretation draws on activity-based travel theory, transport-demand representations, and the recurring native outputs of synthetic mobility studies \citep{AxhausenGarling1992,BowmanBenAkiva2001,McNally2000,Pappalardo2018,Felbermair2020,Kashiyama2024,Kapp2024}. It is an analytical relation rather than a mandatory generation pipeline: methods may generate any one representation directly or connect only a subset of them. Figure~\ref{fig:mobility_representations} summarises this synthesis.

\section{Literature Collection and Review Strategy}
\label{sec:collection}
\label{sec:litcollection}

\begin{figure}[!htbp]
\centering
\resizebox{\linewidth}{!}{%
\begin{tikzpicture}[
    font=\small,
    node distance=7mm,
    box/.style={
        draw,
        rounded corners=2pt,
        align=left,
        inner sep=5pt,
        text width=3.3cm
    },
    flow/.style={
        draw,
        rounded corners=2pt,
        align=center,
        inner sep=6pt,
        text width=11.0cm
    },
    note/.style={
        draw,
        dashed,
        rounded corners=2pt,
        align=left,
        inner sep=5pt,
        text width=11.0cm
    },
    arrow/.style={-Latex, thick}
]

\node[box, fill=gray!8] (mobility) {
\textbf{Mobility terms}\\
human mobility; urban mobility; pedestrian*; traveler*; commuter*
};

\node[box, fill=gray!8, right=5mm of mobility] (generation) {
\textbf{Generation terms}\\
trajectory generation; trajectory synthesis; synthetic trajectory; synthetic mobility; mobility synthesis; activity generation
};

\node[box, fill=gray!8, right=5mm of generation] (methods) {
\textbf{Method terms}\\
GeoAI; GIS; urban computing; deep learning; GAN; VAE; transformer; LLM; agent-based; ABM
};

\node[box, fill=gray!8, right=5mm of methods] (privacy) {
\textbf{Recall terms}\\
differential privacy; privacy-preserving; anonym*
};

\node[draw, rounded corners=2pt, fit=(mobility)(generation)(methods)(privacy),
      inner sep=5pt, label={[font=\bfseries]above:Scopus title--abstract--keyword search}] (searchfit) {};

\node[flow, fill=blue!6, below=12mm of searchfit] (scopus) {
\textbf{Initial database search}\\
Scopus TITLE-ABS-KEY query combining mobility, synthetic-generation, methodological, and recall terms\\
Publication year: 2008--2026\\
\textbf{Records returned: 219}\\
\textbf{After author screening for relevance: 126}
};

\node[flow, fill=blue!6, below=7mm of scopus] (screening) {
\textbf{Title and abstract screening}\\
Retained studies directly related to synthetic human mobility generation, synthetic mobility-relevant populations, activity/trip generation, trajectory generation, people-flow, or aggregate mobility outputs\\
\textbf{Records retained for closer examination: 92}
};

\node[flow, fill=green!7, below=7mm of screening] (expansion) {
\textbf{Review-guided expansion and citation tracing}\\
Used major reviews and representative papers as entry points; conducted backward and forward citation tracing across population synthesis, activity-based simulation, trajectory generation, transformer-based mobility models, and LLM-agentic workflows\\
\textbf{Working corpus after expansion: 121}
};

\node[note, fill=orange!6, below=7mm of expansion] (boundary) {
\textbf{Boundary rule}\\
Adjacent literatures such as next-location prediction, crowd-flow forecasting, dynamic population mapping, map matching, route recommendation, and privacy-preserving mobility release were included only when they directly informed synthetic mobility generation, infilling, artificial mobility construction, or aggregate output concepts.
};

\node[flow, fill=purple!7, below=7mm of boundary] (corpus) {
\textbf{Representative working corpus}\\
Structured review corpus covering core output concepts, methodological families, and practical capability requirements for synthetic human mobility generation
};

\draw[arrow] (searchfit.south) -- (scopus.north);
\draw[arrow] (scopus.south) -- (screening.north);
\draw[arrow] (screening.south) -- (expansion.north);
\draw[arrow] (expansion.south) -- (boundary.north);
\draw[arrow] (boundary.south) -- (corpus.north);

\end{tikzpicture}
}
\caption{Literature collection strategy. The review corpus was assembled through a Scopus title--abstract--keyword search, title and abstract screening, review-guided expansion, and backward and forward citation tracing. The search was designed to capture representative work across synthetic human mobility generation rather than to conduct an exhaustive bibliometric review.}
\label{fig:literature_collection}
\end{figure}
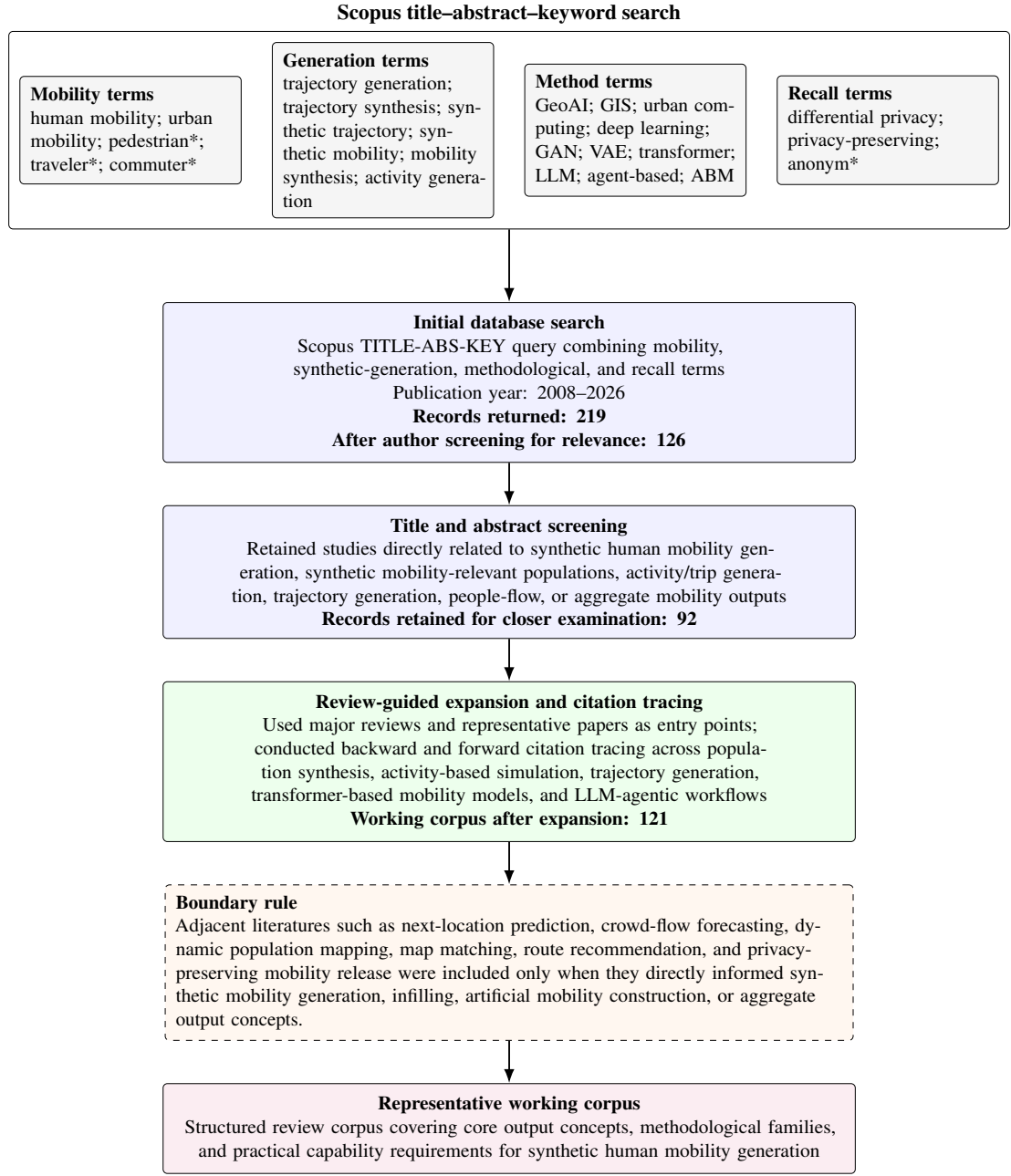

Figure~\ref{fig:literature_collection} summarises the literature collection process. We began with a Scopus search aimed at the core design space of synthetic human mobility generation. The query combined terms related to human and urban mobility, to synthetic or generated mobility data, and to major methodological directions, including trajectory and activity generation, agent-based simulation, deep generative models, transformers, large language models, and privacy-preserving generation. Its generation terms are trajectory- and mobility-synthesis phrases, so the query does not by itself retrieve population-synthesis, activity-based or travel-demand work that does not also use those terms; that literature entered the corpus through the citation-based expansion described below. The purpose of the search was to capture representative work across the field rather than to cover every neighbouring literature exhaustively.

After the initial search, records were screened by title and abstract. We retained studies that made a direct contribution to synthetic human mobility generation or to its immediate design space. This included papers that generated synthetic mobility-related records, constructed synthetic populations linked to mobility, produced daily activity schedules or trip records, generated trajectories, or modelled aggregate mobility outputs as part of synthetic people-flow or mobility generation. We also retained papers that clarified important output forms, methodological families, or evaluation concerns. Several existing reviews were used as entry points to define the surrounding landscape, especially reviews of human mobility modelling, mobile-phone data analysis, trajectory generation, deep learning for human mobility, synthetic urban mobility data, and location privacy \citep{Barbosa2018,Blondel2015,Kong2023,Kapp2024,Luca2021,JiangPrivacy2022}. The corpus was then expanded through backward and forward citation tracing from representative studies, including work on population synthesis, activity-based and agent-based simulation, diary-to-trajectory generation, trajectory-focused generative models, transformer-based mobility models, and LLM-agentic mobility generation \citep{Beckman1996,Pappalardo2018,Felbermair2020,Kashiyama2024,ZhuSynMob2023,Hsu2024,WangUrbanResidents2024,LiMobAgent2024}.

Studies were included when they helped explain what kind of synthetic mobility output is generated, how it is generated, or what assumptions and inputs are required. We therefore kept classic statistical and simulation-based approaches alongside recent neural and LLM-based methods. This choice reflects the aim of the review: to map the methodological landscape of synthetic human mobility generation rather than to focus only on recent deep-learning models. By contrast, papers centred purely on next-location prediction, crowd-flow forecasting, traffic state estimation, map matching, route recommendation, or autonomous-vehicle motion planning were not treated as part of the core corpus unless their modelling logic was explicitly adapted for synthetic generation, infilling, or artificial mobility construction. Dynamic population mapping and aggregate flow estimation were treated as adjacent literatures that help define aggregate mobility outputs, but not as core synthetic mobility generation unless they were directly connected to synthetic people-flow or mobility synthesis. Privacy-preserving studies were included only when they explicitly generated, transformed, or released synthetic mobility records; privacy protection itself is treated in this review as a design and evaluation concern rather than as the primary organising axis.

The Scopus search was executed on 31 July 2026 and returned 219 records; the full query is reproduced in Appendix~\ref{app:scopus}. Author screening for relevance to synthetic mobility generation reduced these to 126 records, and after title and abstract screening, 92 papers were retained for closer examination. Citation-based expansion then brought the working corpus to 121 studies, adding methodological papers, foundational transport and population-synthesis work, review articles, data-resource papers, and recent LLM-agentic studies. The resulting corpus is used as a representative working corpus for a structured review. The reference list does not correspond one-to-one with this corpus. It also includes works reached through the reference lists of review articles, and works identified by tracing the methodological lineage of individual approaches, a number of which predate the search window. It is intended to cover the main output forms and methodological families of synthetic human mobility generation, while recognising that adjacent literatures such as general mobility prediction, traffic forecasting, location privacy, and dynamic population estimation are broader than the scope of this paper.

\section{Methodological Families of Synthetic Human Mobility Generation}
\label{sec:families}

\begin{figure}[!htbp]
\centering
\includegraphics[width=1.\linewidth]{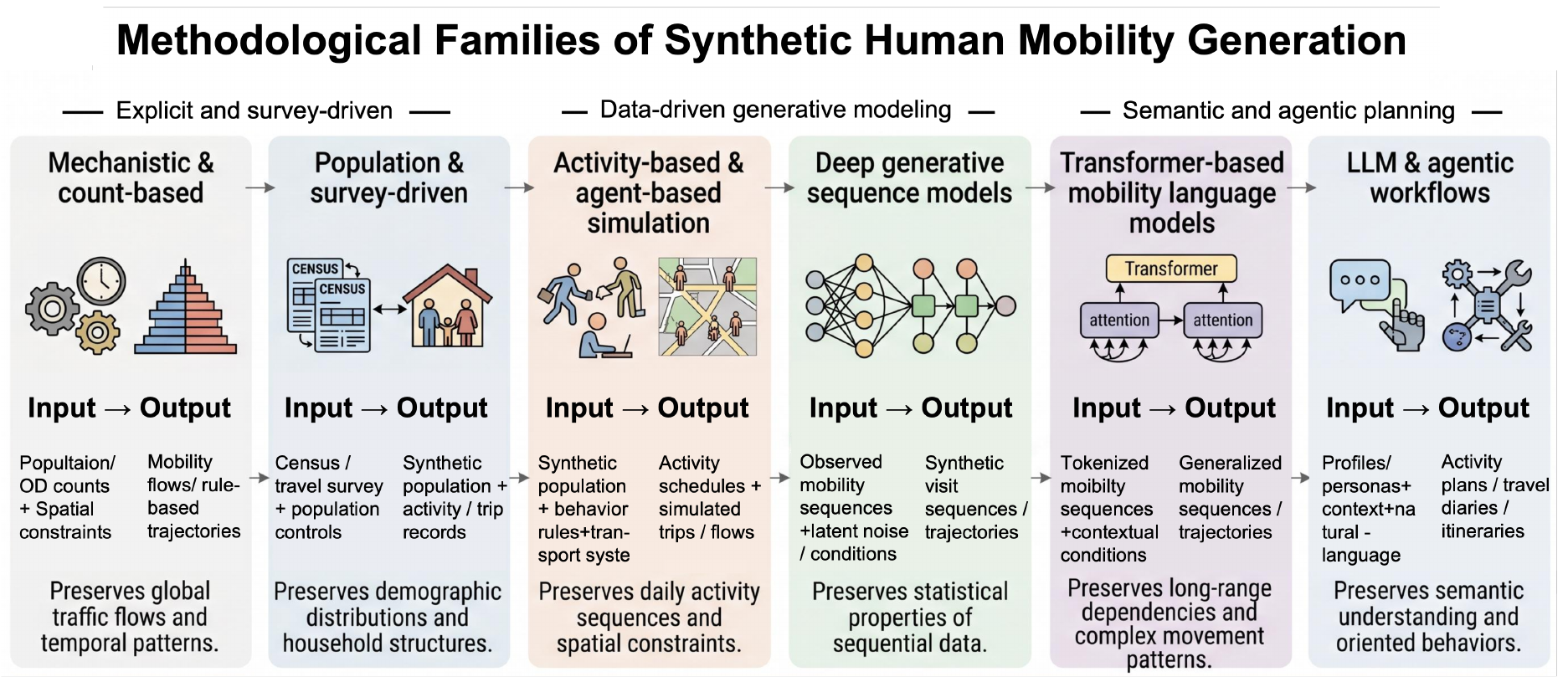}
\caption{Methodological families of synthetic human mobility generation.
The figure summarises the main methodological traditions reviewed in Section~\ref{sec:families}, their typical inputs and native outputs, and conceptual relationships, from explicit and survey-driven approaches to data-driven generative models and semantically richer LLM and agentic workflows. ChatGPT and PicDoc were used to generate icon elements for this figure and to refine its layout; the authors reviewed the content and take full responsibility for it.}
\label{fig:view}
\end{figure}

Synthetic human mobility is generated through several methodological traditions, ranging from mechanistic and survey-driven models to deep generative and language-model-based approaches. These traditions differ not only in model form, but also in what they treat as the primary object of generation, how strongly they rely on prior behavioural structure, how directly they learn from observed data, and which constraints they impose explicitly. Some methods generate mobility through staged pipelines grounded in surveys, synthetic populations, or activity structures. Others learn directly from observed trips, trajectories, or aggregate patterns and generate new samples in the same representation as the training data. More recent approaches extend this further by using transformers, diffusion models, or LLM-based planning to generate mobility with stronger sequence modelling, controllability, or semantic richness. Reviewing these families separately is useful because similar outputs may arise from very different methodological assumptions, while methods with similar labels may differ substantially in native output, semantics, scalability, and downstream requirements. Figure~\ref{fig:view} summarises broad methodological traditions rather than every individual model architecture.

\subsection{Mechanistic and count-based models}

Mechanistic and count-based models provide one of the longest-standing methodological baselines for synthetic human mobility generation. Their common feature is that human mobility is generated from explicit structure rather than learned end-to-end from raw data. This structure may take the form of spatial-interaction laws, trip-distribution rules, exploration--return mechanisms, calibrated behavioural parameters, empirical transition frequencies, or Markovian state transitions \citep{Wilson1970,McNally2000,OrtuzarWillumsen2011,JiangTimeGeo2016,Pappalardo2018,Cornacchia2021,AshbrookStarner2002,Gambs2012,Mathew2012,Liu2024Act2Loc}. Act2Loc is included here for its mechanistic location-assignment component; its activity sequences are learned, and it is discussed as a hybrid method below. These mechanistic models differ substantially in output and scale, but they share a similar logic: synthetic human mobility is produced by combining interpretable rules, calibrated parameters, or local probabilities that can be inspected and controlled.

Three strands are especially relevant. The first is the spatial-interaction and trip-distribution tradition, where aggregate demands, OD flows, or destination choices are generated through gravity-style, entropy-based, or related transport-demand formulations \citep{Wilson1970,Simini2012,McNally2000,OrtuzarWillumsen2011}. The second is mechanistic individual-mobility modelling, where mobility patterns are generated from interpretable assumptions about routines, exploration, preferential return, temporal rhythms, or social-spatial structure. TimeGeo, Ditras, and STS-EPR illustrate this line by combining spatial, temporal, diary-based, or social mechanisms to generate individual mobility profiles or trajectories \citep{JiangTimeGeo2016,Pappalardo2018,Cornacchia2021}. The third is count-based sequential modelling, where movements are sampled from observed frequencies of locations, stays, semantic states, or transitions between discrete states \citep{AshbrookStarner2002,Gambs2012,Baratchi2014,Qiao2018,Anda2021,WangOsaragi2024}. Decision-process formulations extend this strand by treating the daily itinerary as a sequence of rewarded choices recovered from observed traces, whether through Markov decision processes fitted to GPS data \citep{Pang2017} or through context-aware inverse reinforcement learning of activity schedules \citep{Liu2025}. Graph-based inverse reinforcement learning has also been used to generate trajectories where observations are incomplete, drawing on environmental semantics such as building attributes to compensate for sparse data \citep{LinGraphIRL2026}. The last strand provides an important bridge between traditional human mobility modelling and later sequence-based deep neural network models, e.g., recurrent neural networks (RNNs), transformers, or LLMs. Human mobility was already modelled as an ordered sequence of states or transitions before the deep learning and LLM era.

These models can operate on several mobility representations. Spatial-interaction and trip-distribution models most naturally produce aggregate outputs such as OD demand, traffic volume, or flow patterns. Mechanistic individual-mobility models can generate daily mobility profiles, diaries, or trajectories. Count-based sequential models can operate on daily activity schedules, staypoint sequences, trip-like transitions, or discretised trajectories, depending on how the state space is defined \citep{Anda2021,WangOsaragi2024}. This flexibility is one reason the family remains important. The same general machinery---rules, counts, transition probabilities, and calibrated parameters---can be adapted to different output forms. At the same time, these methods usually generate only one part of the broader human mobility representation. Aggregate demand may still require disaggregation, while token or transition sequences may require semantic interpretation, spatial grounding, routing, or aggregation before they become useful for urban analytics.

The main strengths of this family are transparency, data efficiency, and control. Because the generative logic is explicit, these models are often easier to explain or calibrate against known marginals, transition patterns, or demand targets than more complex learned generators \citep{OrtuzarWillumsen2011,JiangTimeGeo2016,Feng2020,Jiang2023}. They are also useful when available data is limited, structured, or already aggregated. In practical review terms, they remain valuable as interpretable baselines, as simple generators of OD demand or daily schedules, and as components that can be embedded within larger synthesis pipelines \citep{Pappalardo2018,Anda2021,Kapp2024}. This distinguishes them from the population-first pipelines discussed next: mechanistic and count-based models do not necessarily begin by constructing a target synthetic population, although they may later be combined with one.

Their limitations follow from the same simplicity. Spatial-interaction models often rely on assumptions about distance decay, independence, or behavioural homogeneity, which can limit their ability to represent individual heterogeneity or behavioural meaning \citep{Wilson1970,OrtuzarWillumsen2011}. Count-based and Markov-style models are usually strongest at local transitions and short-range dependence, but weaker at capturing longer daily structure, cross-scale consistency, or latent behavioural logic. Higher-order and memory-aware studies show that human mobility can contain dependence structures that simple Markov assumptions do not capture reliably \citep{Matamalas2016,Kulkarni2019}. When the state space is coarse, these models may over-simplify mobility; when it is fine, they may become sparse and brittle. They also have limited capacity to represent demographics, household structure, personas, or policy scenarios unless additional modules are added. For this reason, mechanistic and count-based models remain important as interpretable foundations, but they are usually insufficient on their own when the goal is behaviourally meaningful, population-grounded, and differentiated synthetic mobility.

In addition, privacy-related synthetic mobility methods sometimes extend this family when they generate data from privatised counts, marginals, transition matrices, or patterns. DP-WHERE applies differential privacy to human mobility modelling from cellular data, while adaptive n-gram and locally differentially private trajectory synthesis methods use privatised transition or pattern information to generate synthetic trajectories \citep{Mir2013,Gursoy2018,Liu2019,Du2023,Hu2024}. Clustering-based approaches similarly aim to improve the utility of differentially private synthetic trajectories \citep{Yao2022}. Learned generators have been developed for the same purpose, including imitation-learning generators trained without centrally pooling real trajectories \citep{Wang2023} and graph-autoencoder methods that synthesise city-wide mobility from grid-based data \citep{Netzler2024}. A related line represents individual mobility as a graph rather than a sequence, generating interdependent attributes together with semantic location types such as home, transit stop, and workplace \citep{BostanipourMobilityGraphs2026}. Consistent with the scope stated in Section~\ref{sec:collection}, these are treated here as a design and evaluation concern rather than as a separate methodological family. The central trade-off is that noise added to sparse or low-frequency states can distort temporal coherence, semantic structure, and trajectory plausibility once synthetic mobility is reconstructed from privatised statistics.

\subsection{Population synthesis and survey-driven mobility pipelines}
\label{sec:popsyn}

Population synthesis and survey-driven mobility pipelines generate mobility by starting from a representation of people rather than from isolated traces. This family builds on the three population concepts defined in Section~\ref{sec:scope}: static synthetic population, dynamic or de facto population distribution, and synthetic people-flow or mobility population. Synthetic human mobility sits between these traditions. It starts from synthetic residents or population priors, but its usefulness is often judged by whether it can reproduce dynamic population distributions, flows, and activity patterns across the urban day.

The static synthetic population tradition provides the foundation for this family. Early work in population synthesis was developed to construct disaggregate households and individuals for microsimulation and activity-based travel demand models when complete individual-level census records are unavailable because of privacy, cost, or sampling constraints \citep{Beckman1996,GuoBhat2007,muller2010population}. Subsequent work extended this tradition by improving sample-based synthesis, addressing cases where microdata are unavailable, and reviewing synthetic population construction for social and agent-based simulations \citep{BarthelemyToint2013,Chapuis2022}. In urban modelling, synthetic populations became a core infrastructure for integrated land-use, transport, and policy simulation, as illustrated by UrbanSim and later spatial microsimulation frameworks such as UrbanPop \citep{Waddell2002,UrbanPop2023}. These studies make clear that population synthesis is not merely demographic annotation. It constructs the agent base from which activity participation, travel demand, and policy response can be simulated.

The same logic appears in survey-driven synthetic mobility pipelines. Individuals or households are first synthesised or sampled, assigned demographic and household attributes, and then linked to activity schedules, trip chains, destinations, modes, or trajectories. \citet{Felbermair2020}, for example, generate a synthetic population with activity chains as input for agent-based transport modelling, using statistical raster census data and limited survey information to assign realistic travel plans to agents. A nationwide synthetic human mobility dataset has been constructed by combining limited travel surveys with open statistical data, showing how survey information and open data can be used to scale synthetic mobility beyond a local case study \citep{Kashiyama2024}. Recent work has also begun to connect population synthesis with deep learning for generating individual spatiotemporal daily activity schedules \citep{Lu2026}. Related pipelines start from population-level mobility models in order to simulate shareable mobility data \citep{Smolak2020}, or use large-scale mobile phone positioning data to construct individualised pseudo-personal mobility records for a study population \citep{Li2025a}. Related Japanese work has developed household-level synthetic populations and distribution systems for large-scale social simulation, including agent-based household micro-datasets and Japanese synthetic population data with protection levels \citep{Sugiki2012,MurataSyntheticPopulation}. These examples show how synthetic population becomes synthetic mobility when population attributes are translated into activities, trips, or people-flow records.

A separate but closely related line of work concerns dynamic population distributions. Unlike static synthetic populations, these outputs do not necessarily generate individual agents. Instead, they estimate or represent where people are located at different times of day. Mobile phone data, administrative data fusion, and gridded population modelling have been used to estimate dynamic, ambient, or time-specific population distributions at national, regional, and metropolitan scales \citep{Deville2014,Martin2015,Bergroth2022,Khodabandelou2019,Yang2021}. Global and national population products such as WorldPop and LandScan also illustrate the importance of moving beyond resident population toward spatially explicit population presence and exposure \citep{Tatem2017WorldPop,Dobson2000LandScan}. In Japan, Mobile Spatial Statistics estimate actual population distributions from mobile terminal network data, while household transition models extend synthetic household data toward future population and household distributions \citep{Terada2013MSS,Kajiwara2022}. In the UK, Population 24/7 and UK government time-specific population statistics show how time-specific population modelling can support planning, emergency response, and service delivery \citep{Population247NRT2021,ONS2023TimeOfDay,ONS2025PMD}. These studies are not always synthetic mobility generators, but they provide important calibration and validation targets for synthetic mobility.

This distinction is important for urban analytics. Static synthetic population answers the question of \textit{who} is represented. Synthetic people-flow answers how those people are translated into activities, trips, and movements. Dynamic population distribution answers whether the resulting population presence across the city is plausible over time. A synthetic mobility method may generate realistic individuals but fail to reproduce daytime population distribution; it may match aggregate population flows but lack plausible household routines; or it may generate trajectories without any link to a target population. For this reason, population-grounded synthetic mobility should not only match residential population distributions. It should also reproduce dynamic de facto population patterns, OD structures, and temporal demand rhythms when aggregated.

The main strengths of population synthesis and survey-driven mobility pipelines are therefore behavioural interpretability, demographic heterogeneity, and policy relevance. Because the synthetic population carries explicit demographic and household information, it becomes easier to study how mobility differs across age groups, income levels, occupations, household types, car-ownership categories, or residential contexts \citep{Felbermair2020,Kashiyama2024,Wu2022GBSyntheticPopulation}. This also makes the family well suited to scenario analysis. Land-use change, infrastructure interventions, service disruptions, demographic aging, household transition, or policy shifts can be introduced by altering opportunities, constraints, or behavioural assumptions for different population segments.

Their limitations are also clear. These pipelines depend heavily on survey structure, census variables, coding schemes, and synthetic-population assumptions. They may inherit survey simplifications, outdated behavioural priors, or mismatches between residential population and de facto population patterns \citep{StopherGreaves2007,Kapp2024}. Detailed trajectories are often not native outputs; they usually require additional place assignment, routing, simulation, or refinement after activities and trips have already been generated \citep{Felbermair2020,Kashiyama2024}. For this reason, population synthesis and survey-driven mobility pipelines are especially strong when the goal is to generate semantically rich and policy-relevant synthetic mobility, but they need to be coupled with trajectory grounding, dynamic population calibration, and aggregate validation to support realistic large-scale mobility generation.

\subsection{Activity-based and agent-based simulation}

Activity-based and agent-based simulation treat mobility as the outcome of explicit behavioural processes unfolding over time. Rather than generating isolated trips or trajectories directly from observed counts or traces, this family models how individuals or households make activity, destination, timing, and mode decisions under constraints, and how those decisions translate into movement once they are executed in a simulated environment. This view is closely related to activity-based travel demand modelling, where activity participation and scheduling decisions are treated as central components of travel behaviour \citep{Nayak2023}. It is also aligned with agent-based simulation work that represents individuals as agents embedded in social, spatial, and behavioural contexts \citep{Ronald2012,Zufle2023}. In this setting, synthetic mobility is not only sampled from a distribution. It is produced through a process that attempts to represent how daily travel arises from schedules, preferences, constraints, and interactions.

This family is closely related to population-first synthesis pipelines, but it is more explicit about execution and behavioural dynamics. In many cases, a synthetic population is first constructed and assigned attributes, after which agents are given activity plans, daily schedules, or decision rules. The distinctive step comes later: these plans are not treated as final outputs, but are executed, revised, or iteratively adjusted through simulation. The raster-census pipeline described in Section~\ref{sec:popsyn} is one example \citep{Felbermair2020}. \citet{Ronald2012} demonstrate agent-based simulation of social activity generation and scheduling with social networks, while \citet{Zufle2023} model urban life through agents, people, places, needs, and social structures. These examples show that activity-based and agent-based simulation is not only about producing trips, but also about representing the behavioural and social processes from which trips arise.

Execution is the feature that most clearly distinguishes this family from survey-driven synthesis alone. Routes may be assigned on transport networks, congestion may emerge through network loading, and activity timing or mode choice may respond to simulated conditions. A reinforcement-learning-based framework has been developed for simulating people mass movement from GPS data \citep{Pang2020}, and agent-based modelling has been combined with particle filtering to follow mobile spatial statistics \citep{Cai2021}. Simulation-based trajectory learning has also been used to generate or refine spatiotemporal trajectories under controlled system conditions \citep{Glake2022}. Alongside MATSim, open-source microscopic traffic simulators such as SUMO provide the network-loading and route-choice machinery on which many such experiments depend \citep{Behrisch2011}. These works illustrate why this family is especially important when mobility must remain feasible not only at the level of isolated individuals, but also under system-level interaction, feedback, or observation constraints.

The native outputs of this family are often richer than a single representational layer. Depending on the implementation, activity-based and agent-based simulation may produce activity schedules, trip chains, route choices, link flows, executed trajectories, and network performance indicators within the same pipeline \citep{Horni2016,Pang2020,Cai2021}. This makes the family especially relevant for urban analytics and policy evaluation. Interventions such as pricing, service changes, infrastructure expansion, disruptions, or land-use shifts can be introduced into the simulated environment, and their consequences can then be traced through both individual behaviour and aggregate system outcomes. In contrast to purely trace-driven generators, these methods often preserve the link between who moves, why they move, and how the transport system responds.

A major strength of this family is its ability to connect micro-level decision logic with macro-level urban consequences. Demographic heterogeneity, household structure, accessibility constraints, and service supply can all be represented explicitly, making it possible to study distributional effects, behavioural adaptation, and scenario-dependent change \citep{Zufle2023,Kashiyama2024}. Because execution takes place in a constrained environment, the resulting mobility is often stronger in feasibility than methods that generate trips or trajectories without explicit network or system feedback. This also makes the family well suited to applications where congestion, route competition, timetable interaction, or capacity limits matter directly \citep{Pang2020,Cai2021,Glake2022}.

Hybrid approaches also connect behaviourally structured representations
with trajectory generation. Act2Loc first generates individual activity
sequences using machine-learning methods and then applies mechanistic
models to assign activity locations, thereby translating activity-level
patterns into synthetic trajectories \citep{Liu2024Act2Loc}. It therefore
illustrates how data-driven sequence generation can be combined with
mechanistic spatial grounding, rather than treating trajectory synthesis
as a purely end-to-end statistical task.

The limitations of this family are equally important. Activity-based and agent-based simulation typically requires many assumptions, multiple calibrated submodels, and substantial external inputs, including population data, behavioural parameters, network data, and often activity or trip distributions from surveys \citep{Nayak2023,Kapp2024}. Results may therefore depend heavily on design choices that are difficult to validate independently. In addition, although these methods are strong at preserving behavioural logic and feasibility, they do not necessarily reproduce the fine-grained statistical patterns of observed trajectories unless they are carefully calibrated or combined with trace-based refinement \citep{Glake2022,Kapp2024}. Computational cost can also become substantial when simulation is scaled to large populations or when iterative replanning is required. For this reason, activity-based and agent-based simulation remains one of the strongest families for semantically rich, policy-relevant, and system-aware synthetic mobility, but it is not always the most efficient route to high-fidelity trace generation.

\subsection{Deep generative and trajectory-based models}

Deep generative sequence models treat synthetic human mobility as a data-driven generation problem. Instead of specifying human mobility through explicit behavioural rules, survey-based schedules, or simulation logic, this family learns statistical regularities directly from observed mobility data and samples new sequences from the learned distribution. The observed data may be represented as activity tokens, trips, discretised locations, link sequences, or continuous coordinates, depending on the application. What unifies this family is not a single architecture, but a shift in modelling logic: the generator is trained to reproduce patterns in a training corpus rather than to execute an explicitly specified behavioural process \citep{Goodfellow2014,YuSeqGAN2017,Kulkarni2017,Feng2020,rao2020lstm,Berke2022,Kapp2024,Luca2021}.

This family became prominent as large mobility datasets grew more available and deep sequence modelling matured. Earlier count-based and Markov-style approaches already treated mobility as an ordered sequence, but deep generative models extended this idea by learning richer spatial, temporal, and sequential dependencies from data \citep{AshbrookStarner2002,Gambs2012,Mathew2012, KingmaWelling2014,Kulkarni2017,Feng2020}. In synthetic mobility research, this often meant moving from transition-count sampling to recurrent architectures, adversarial training, latent-variable models, or neural point-process formulations. Depending on the training representation, these models may generate activity-like token sequences, trip-like event sequences, discretised mobility traces, or continuous trajectories. A minority emit activity semantics directly rather than as a conditioning signal, generating activity trajectories by modelling the spatiotemporal dynamics that link where a person is to what they are doing \citep{YuanActivityTraj2022}. In practice, however, this family has been especially influential for trajectory-like outputs, because high-frequency GPS and location-trace data provide abundant training material at that level \citep{Wang2021,Jiang2023,Zhang2023,Rao2023,Long2023}.

The native outputs of this family are usually synthetic sequences in the same representation as the training data. This is an important distinction from population-first or activity-based pipelines. If the model is trained on discretised grid or POI sequences, it typically generates new sequences in that same tokenised space. If it is trained on continuous trajectories or embedded mobility traces, it usually outputs synthetic traces in comparable form. Context labels such as time, user attributes, mode, or purpose may be added as conditioning variables, but they are not always part of the native output itself \citep{Rao2023,Jia2024,Cao2025}. For this reason, deep generative sequence models are often strongest when the research goal is to reproduce the statistical structure of observed mobility rather than to construct a fully interpretable behavioural pipeline from population attributes to urban-system outcomes.

A recent specialisation within this family is diffusion-based trajectory generation. Models such as DiffTraj, profile-guided latent diffusion, topology-constrained diffusion, and collaborative-noise diffusion generate trajectories through iterative denoising or refinement, often improving local geometric fidelity and controllability \citep{Chu2023,chu2024simulating,ZhuDiffTraj2023,Song2024,Zhu2024,Zhang2025,Luo2026,Zou2026}. Two recent variants extend the conditioning signal beyond the trajectory itself: one guides denoising with dynamic population distribution so that generated movement remains consistent with where people actually are at each time \citep{LongDynPopDiff2026}, and a two-stage design first recovers latent movement sequences by diffusion and then decodes fine-grained check-ins with their points of interest and timestamps \citep{XuGeoGen2026}. Their role in this review is therefore not to define a separate methodological family, but to illustrate how trajectory-focused deep generative models are being extended toward smoother, more constrained, and more controllable traces. Like other trajectory-centred generators, however, they usually require additional semantic grounding and population-level calibration when the goal is practical synthetic mobility rather than trajectory imitation alone.

The strengths of this family lie in flexibility and expressive power. Compared with mechanistic or count-based methods, deep generative models can capture more complex dependencies among location, timing, and sequence structure, especially when the training corpus is large and the representation is well chosen. \citet{Kulkarni2017} use recurrent neural networks to generate synthetic mobility traffic, while \citet{Feng2020} propose a generative adversarial framework for learning to simulate human mobility. \citet{Berke2022} use RNNs to generate synthetic mobility data for a realistic population while balancing utility and privacy. GAN-based trajectory generators and conditional adversarial models further extend this data-driven logic to continuous or modality-aware trajectory synthesis \citep{Wang2021,Jiang2023,Zhang2023,Rao2023,Jia2024,Cao2025,Crivellari2025}. Comparative work has assessed how far such generators reproduce mobility characteristics rather than merely matching them superficially \citep{kulkarni2018generative}, and further variants separate stopping points from moving points in a two-stage design \citep{Gong2023}, retrieve and recombine observed fragments instead of sampling from scratch \citep{Huang2023}, or learn continuous latent fields with implicit neural representations rather than discrete sequences \citep{Tenzer2024}. At city scale, deep generative networks have been used to couple individual movement, population flows, and urban morphology within a single model \citep{Yuan2025}.

Two developments push this family beyond pure imitation. Conditional formulations generate trajectories under contextual signals such as social events rather than reproducing an unconditional data distribution \citep{Deng2025}, and semantic-numerical frameworks aim at controllable generation by combining fine-grained spatiotemporal dependencies with higher-level semantic context, so far mainly for vehicle traffic \citep{Lakmal2025}. Both are attempts to move generation from statistical fidelity toward the behavioural specificity discussed in Section~\ref{sec:mpa}.

Their limitations are equally important. These models often preserve what is statistically salient in the training data without preserving what is behaviourally meaningful. A generated sequence may look plausible at the level of local motion or short-range temporal structure while still failing to satisfy route feasibility, activity logic, or macro-level urban regularities once many samples are combined \citep{Kapp2024,Kong2023}. This is especially important in urban analytics, where a synthetic mobility model is often expected not only to mimic observed traces, but also to remain interpretable, controllable, and consistent with known population patterns. For this reason, deep generative sequence models are frequently coupled with additional modules for map matching, routing, constraint enforcement, or macro calibration when they are used for realistic urban-scale synthesis.

A further limitation is that semantics and heterogeneity are not always native strengths of this family. Demographic attributes, household roles, activity meaning, and scenario conditions can sometimes be added as conditioning inputs, but they are often external controls rather than structurally central parts of the generator. This limitation helps explain why later methodological strands---especially diffusion-based trajectory generators, transformer-based mobility language models, and LLM-oriented planning methods---have attracted attention. These newer families seek, in different ways, to improve geometric fidelity, sequence control, long-range dependence, or semantic interpretability. Deep generative sequence models therefore occupy an important middle position in the evolution of synthetic human mobility: they move beyond explicit rules and shallow transition statistics, but they do not by themselves resolve the broader challenge of generating semantically rich, feasible, and population-consistent mobility.

\subsection{Transformer-based mobility language models}

Transformer-based mobility language models treat mobility as a token sequence and generate synthetic mobility through next-token prediction, conditional completion, or infilling over discretised representations of space and time. In this family, mobility is cast into a language-like form: locations, links, grid cells, POIs, time bins, and sometimes control variables are encoded as tokens, and the model learns to generate coherent sequences under this tokenisation scheme. What distinguishes this family is therefore not only the use of the transformer architecture, but the modelling assumption that mobility can be expressed as an ordered symbolic sequence with learnable long-range dependencies \citep{Hsu2024,Kobayashi2023,Haydari20261681}.

This family emerged as a natural extension of earlier sequence models. Markov and recurrent approaches already treated mobility as ordered states, but transformer-based models strengthened the ability to capture longer-range dependencies, contextual constraints, and partially observed sequence structure \citep{Gambs2012,LiuST-RNN2016,FengDeepMove2018,Hsu2024}. Masked-language-model pretraining has been applied to mobility sequences in the same spirit, using feature enrichment and data augmentation to predict individual movement \citep{Yasuda2025}, and hybrid designs combine transformer attention with graph convolution over the spatial structure \citep{CorriasTransformerGCN2023}. An early demonstration that language-model machinery transfers to this setting reframed mobility forecasting as a sentence-completion problem using a pretrained language foundation model \citep{XueLangFM2022}. This is especially relevant in synthetic mobility, where generation may depend not only on the immediately previous state, but also on earlier routine structure, time context, or constraints supplied at generation time. Compared with earlier recurrent models, transformer-based approaches are often more flexible in conditional generation, partial completion, and controllable sequence synthesis.

A parallel development concerns urban foundation models that pretrain on spatiotemporal data rather than on individual mobility sequences. GPT-ST introduces generative pretraining for spatiotemporal graph neural networks, and UrbanGPT aligns a spatiotemporal dependency encoder with instruction tuning so that a single model can forecast across urban tasks, including in zero-shot settings \citep{LiGPTST2023,LiUrbanGPT2024}. These models are forecasters rather than generators of individual mobility, and are included here for the same reason as earlier aggregate models: they show that the tokenisation and pretraining logic of this family is being applied at urban scale, which is the direction from which population-level grounding is most likely to arrive. A mobility-native counterpart abstracts each geographic area as a token and pretrains on large-scale location-based service traces covering a substantial share of a city's population, producing a foundation model of mobility itself rather than of spatiotemporal signals in general \citep{WuPretrainedMobility2026}.

The native outputs of this family are usually token sequences rather than direct continuous trajectories. These tokens may represent locations, spatial cells, links, visits, or temporally indexed mobility states, and they may be generated from scratch, conditioned on control signals, or used to fill missing parts of a partially specified sequence. Coordinates, point traces, or route-like outputs are often derived only after token decoding, interpolation, smoothing, or feasibility checks. This gives the family a distinctive position within synthetic mobility generation. Compared with diffusion-based models, transformer-based mobility language models are usually less focused on fine geometric continuity. Compared with activity-based and agent-based approaches, they are less explicitly grounded in behavioural process models. Their strength lies instead in sequence-level regularity, controllable completion, and flexible symbolic representation.

This intermediate position is clear in recent work. Generative pretraining has been applied directly to daily trajectories, with GPT-2 used to generate individual daily movement sequences \citep{MizunoGPT2Traj2022} and GPT-style models entered into open mobility-prediction challenges \citep{SolatorioGeoFormer2023}. Transformer models have also been trained to learn daily mobility structure explicitly \citep{WangOsaragiTransformer2024}, and cross-city designs transfer a mobility transformer to cities unseen during training, which matters directly for generating mobility where no local trace data exist \citep{WangCOLA2024}. \citet{Hsu2024} propose TrajGPT, a transformer-based multitask spatiotemporal generative model that frames controlled trajectory generation as a text-infilling problem. \citet{Kobayashi2023} model and generate human mobility trajectories using a transformer with day encoding, while \citet{Haydari20261681} develop MobilityGPT for enhanced human mobility modelling with a GPT architecture. These approaches show how transformer-based mobility models can support controllable or context-aware generation without necessarily producing fully semantic activity or mode outputs as native fields. They are therefore closer to mobility-specific sequence generation than to general-purpose LLM planning.

The strengths of this family lie in long-range sequence modelling, flexible conditioning, and controllable generation in token space. These properties make transformer-based mobility language models well suited to tasks such as partial-sequence completion, constrained generation, or synthetic trace generation when the desired output can be expressed as a structured token sequence. They also provide an important bridge between earlier deep sequence models and later LLM-oriented approaches: they inherit the token-based generative logic of language modelling, but remain centred on mobility-specific corpora and mobility-specific sequence formats rather than on open-ended text reasoning.

At the same time, their limitations are similar to those of other sequence-based trajectory generators. Token-level plausibility does not guarantee route feasibility, semantic interpretability, or correct macro-level urban patterns. Without additional grounding and calibration, a model may generate sequences that are locally coherent but still misaligned with network constraints, population distributions, or aggregate demand. For this reason, transformer-based mobility language models occupy an important intermediate position in the methodological landscape. They move beyond earlier recurrent sequence generators by offering stronger long-range dependency modelling and more flexible control, yet they stop short of the richer semantic planning that characterises LLM and agentic workflows.

\subsection{Large language models and agentic workflows}

Large language models and agentic workflows treat synthetic human mobility as a planning and reasoning problem rather than only as a sequence-generation problem. In this family, the model is asked to produce mobility through semantically meaningful intermediate steps, such as daily plans, travel diaries, itineraries, daily activity schedules, or stay--trip structures, often conditioned on persona, context, prompts, and external information sources. What distinguishes this family is not simply the use of a large language model. It is the shift in generative logic. Instead of learning only from mobility traces in a fixed representation, these methods rely on instruction following, contextual grounding, retrieval, tool use, or multi-step reasoning to generate mobility that is interpretable and controllable at the behavioural level \citep{Zhang2024,WangUrbanResidents2024,LiMobAgent2024,Li2025,Liu2026}. The general capabilities this family draws on---instruction following, in-context learning, and agentic tool use---are surveyed elsewhere \citep{ZhaoLLMSurvey2023,DongICL2023,XiAgentSurvey2025}, and their uptake across urban research has itself been reviewed \citep{XiaUrbanLLM2025}. The immediate precursor is \citet{Park2023}, whose generative agents maintain memory, reflection, and planning to produce believable daily behaviour in a simulated town; the mobility-specific frameworks below adapt that architecture to real urban space.

This family is most naturally associated with the semantic and planning side of synthetic mobility generation. A typical workflow begins with a synthetic person, household, or persona profile, and then generates daily intentions, activities, destinations, or travel episodes before any detailed trajectory exists. \citet{WangUrbanResidents2024} introduce an LLM-agent framework for personal mobility generation that aligns LLMs with real-world human mobility data and uses retrieval-augmented activity generation. \citet{LiMobAgent2024} propose MobAgent for travel diary generation using LLM agents and individual profiles. MobGLM has been introduced for synthetic human mobility generation \citep{Zhang2024}, while \citet{Li2025} propose Geo-Llama for constraint-aware mobility trajectory generation using an LLM-inspired framework. Modular agent frameworks assemble these capabilities into pipelines for full trajectory simulation \citep{JuTrajLLM2025}, and the cognitive-agent framework discussed above is developed further into a complete generative-agent transport simulation \citep{Liu2026,LiuGATSim2026}. A related strand asks whether an LLM can act directly as a mobility predictor rather than a planner, prompting the model with a user's recent stays to infer the next location \citep{WangLLMMob2023,ChenInteractiveNextLoc2025}, and more recent work moves from prompting toward explicitly foundation-model-based generation of mobility trajectories \citep{Li2025b}. Agentic formulations have also been applied to the narrower decision of day-to-day route choice \citep{WangAgenticRoute2025}. A further variant returns the output to the survey instrument itself: agents built from a synthetic population and persona-rich prompts answer a simulated household travel survey, and the resulting synthetic responses are checked against empirical survey distributions \citep{SalvadorLLMAgent2026}. This closes a loop back to the survey-driven pipelines of Section~\ref{sec:popsyn}, since the generated object is the very data source those pipelines depend on. Outside human mobility, the same language-based logic has been used to generate vehicle interaction data for autonomous driving, translating textual descriptions of interactions into driving behaviours and then into trajectories \citep{YangTrajectoryLLM2025}; that work lies outside this review's scope as defined in Section~\ref{sec:scope}, but it indicates how far the text-to-trajectory formulation is being carried. These works show how LLM-based systems can generate or condition higher-level mobility structure.

A major strength of this family is its ability to represent differentiated behaviour in a relatively natural and controllable way. Demographic attributes, preferences, constraints, household roles, and scenario conditions can be expressed directly in prompts, personas, or structured profiles. This makes the family especially attractive for urban analytics applications that require more than geometric realism. Policy evaluation, service redesign, accessibility analysis, disruption response, and scenario-based simulation often depend on who is moving, why they move, and how different groups may respond under changed conditions. LLM-based planning methods are well suited to this kind of variation because they can generate semantically rich daily routines and condition them on contextual narratives or structured inputs \citep{WangUrbanResidents2024,LiMobAgent2024,Liu2026}.

This family also extends the idea of upstream conditioning beyond traditional demographic fields. In many earlier approaches, heterogeneity is introduced through age, gender, occupation, income, household composition, or car ownership. LLM-based methods can incorporate these same attributes, but often go further by using personas that combine demographic information with preferences, constraints, routines, and mobility style. This broader use of personas is consistent with recent work on persona-driven synthetic data and LLM-based simulation \citep{GePersonaHub2025}, where persona design can improve controllability but may also introduce bias, stereotyping, or limited diversity \citep{LutzPersona2025,HuPersona2024,YuSyntheticBias2023}. For synthetic mobility, persona-based conditioning should therefore be understood as a richer semantic layer built on top of synthetic population attributes rather than as a replacement for them.

The native outputs of LLM and agentic workflows are usually not full trajectories. More often, they are plans, itineraries, daily activity schedules, destination choices, travel diaries, or semantically structured stay--trip descriptions that later require grounding, route generation, timing adjustment, and population-scale calibration. This is an important distinction from both deep generative trajectory generators and mobility-specific transformer language models. Those families often generate traces or token sequences directly. LLM and agentic workflows typically generate higher-level mobility structure first and rely on additional modules to transform it into feasible movement in geographic space. Their strength lies in semantic richness and controllability, not in native geometric fidelity.

The limitations of this family are substantial. Rich semantics do not guarantee realistic behaviour. Generated plans may sound plausible while still being inconsistent with travel-time budgets, network structure, service availability, or known urban regularities once grounded. Diversity can also be a problem. This risk is greatest where the persona space is narrow or textually generic \citep{LutzPersona2025}. In addition, these methods often lack the empirical anchoring of survey-based or trace-based models unless they are carefully coupled with retrieval, calibration, or external data sources. For this reason, LLM and agentic workflows are powerful as high-level generators of behavioural structure, but they are rarely sufficient on their own when the goal is realistic, large-scale synthetic mobility.

This family is best understood as a recent extension of the field toward richer semantic planning and more explicit behavioural differentiation. It is especially valuable when synthetic mobility must remain interpretable, scenario-sensitive, and responsive to heterogeneous users or contexts. At the same time, it makes the separation between semantic generation and spatial execution more visible than ever. A plausible daily plan is not yet a plausible mobility trace, and a set of plausible individual plans is not yet a realistic city. This is why the family is likely to remain closely tied to the downstream steps noted above. They push the field toward stronger semantic realism, but they do not remove the need to connect that realism back to trip structure, trajectories, and urban-scale consequences.

\section{Meaning, Population, and Autonomy: Methodological Requirements for Practical Synthetic Mobility}
\label{sec:mpa}

A practical synthetic human mobility method should be assessed on more than the plausibility of individual samples. Depending on the intended application, users may need outputs that retain interpretable behavioural information, represent a specified urban population rather than an uncalibrated collection of traces, and generate mobility for newly synthesised individuals rather than only extending the observed histories of known users. These needs motivate three methodological requirements: \textit{behavioural meaning}, \textit{population grounding and scale}, and \textit{generation autonomy}. We denote these requirements by \textbf{M}, \textbf{P}, and \textbf{A}, respectively. They are treated here as capability dimensions rather than as model families or output categories. Within each dimension the levels are ordered, with two stated exceptions: the two facets of behavioural meaning are not ordered against each other, and P3 records an adjacent capability rather than a stronger form of P2 (Section~\ref{sec:limitations}).

\begin{figure}[!htbp]
\centering
\includegraphics[width=1.\linewidth]{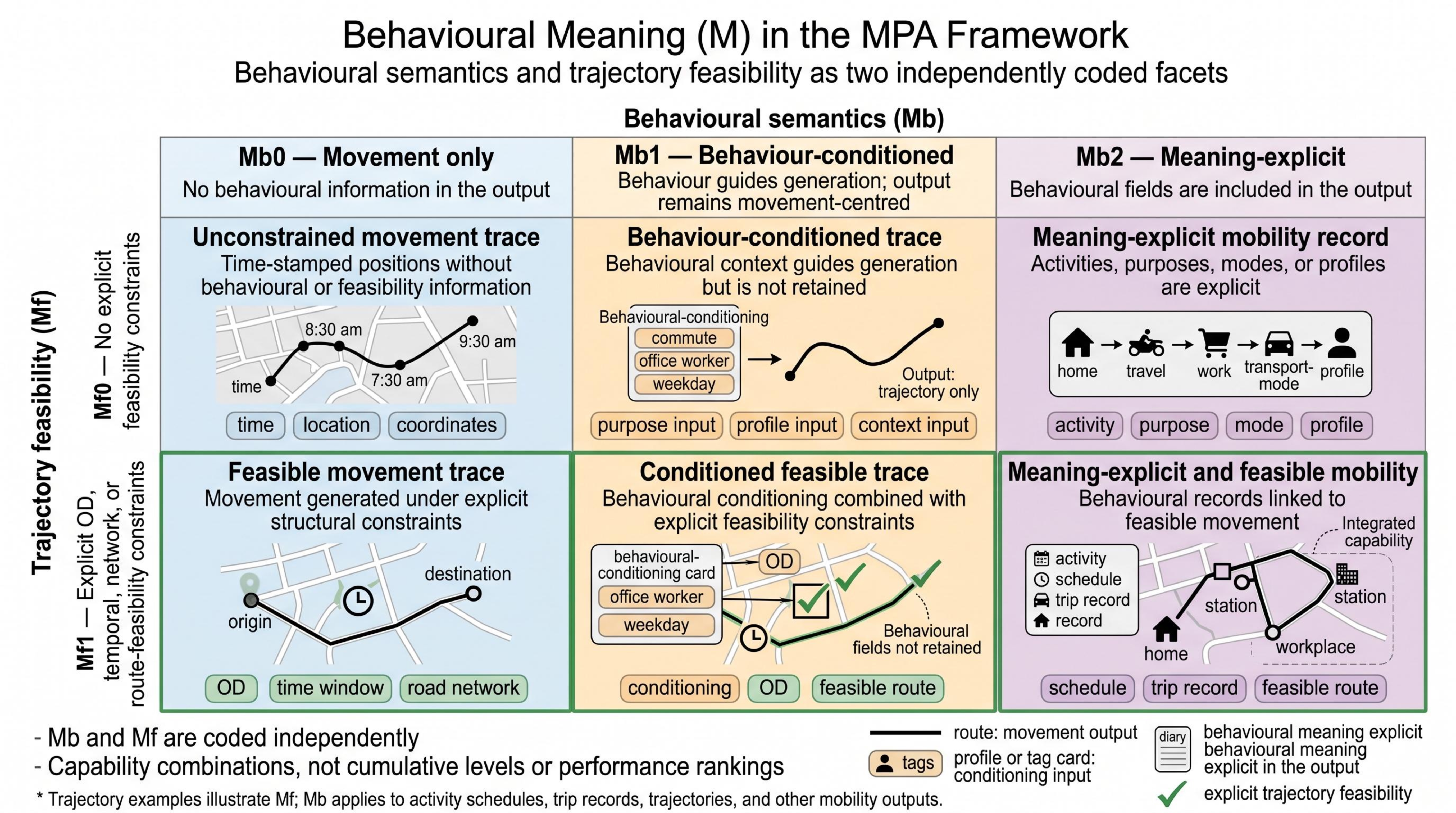}
\caption{Behavioural meaning in the Meaning--Population--Autonomy framework, shown as two independently coded facets. Behavioural semantics (M\textsubscript{b}) runs across the columns: at M\textsubscript{b}0 the output is movement only, at M\textsubscript{b}1 behavioural information guides generation but is not retained in the generated record, and at M\textsubscript{b}2 behavioural fields are explicit in the output. Trajectory feasibility (M\textsubscript{f}) runs down the rows: at M\textsubscript{f}0 generation is not subject to explicit feasibility constraints, and at M\textsubscript{f}1 it is. The six cells are capability combinations rather than cumulative levels, and neither facet is a ranking of model quality; the trajectory examples illustrate M\textsubscript{f}, while M\textsubscript{b} applies equally to activity schedules, trip records and other mobility outputs. ChatGPT and PicDoc were used to generate icon elements for this figure and to refine its layout; the authors reviewed the content and take full responsibility for it.}
\label{fig:behaviour_meaning}
\end{figure}

The purpose of these dimensions is to clarify what kind of problem a method actually solves. A trajectory generator that depends on a partial user history is not equivalent to a generator that creates mobility for a new synthetic population. A semantically rich activity planner is not equivalent to a diffusion model that produces fine-grained traces~\citep{ZhuDiffTraj2023,Song2024,Zhu2024}. A method that matches aggregate OD demand is not equivalent to one that produces realistic individual routines. Table~\ref{tab:mpa_dimensions} provides the operational definitions of all three capability dimensions. Because the distinction between behaviour-conditioned generation (M\textsubscript{b}1) and behaviour-explicit output (M\textsubscript{b}2) is less intuitive than the corresponding population and autonomy levels, Figure~\ref{fig:behaviour_meaning} illustrates the behavioural Meaning dimension in greater detail. The three dimensions are subsequently considered together when positioning representative methods in the MPA capability space. \ref{app:mpa} applies the same criteria to every generation method reviewed in Section~\ref{sec:families}, reporting its native output and its level on each dimension.

\begin{table}[!ht]
\centering
\small
\caption{Operational definitions of the Meaning--Population--Autonomy capability levels. Higher levels indicate stronger capability along the corresponding dimension, rather than higher overall model quality, except where noted for P3 in Section~\ref{sec:limitations}. Behavioural meaning is recorded on two facets, behavioural semantics (M\textsubscript{b}) and trajectory feasibility (M\textsubscript{f}), which are scored separately and are not ordered against each other.}
\label{tab:mpa_dimensions}
\begin{tabular}{p{0.25\linewidth} p{0.10\linewidth} p{0.60\linewidth}}
\toprule
\textbf{Dimension} & \textbf{Level} & \textbf{Operational criterion} \\
\midrule
\textbf{Behavioural semantics (M\textsubscript{b})}
& M\textsubscript{b}0 & Movement-only output, with no explicit behavioural information. \\
& M\textsubscript{b}1 & Behavioural information guides generation but is not included in the generated output. \\
& M\textsubscript{b}2 & Behavioural information is explicitly included in the generated output. \\
\addlinespace
\textbf{Trajectory feasibility (M\textsubscript{f})}
& M\textsubscript{f}0 & Generation is not subject to explicit network, routing, or physical feasibility constraints. \\
& M\textsubscript{f}1 & Generation is subject to explicit network, routing, or physical feasibility constraints. \\

\midrule
\textbf{Population grounding and scale (P)}
& P0 & Individual samples without explicit target-population grounding. \\
& P1 & Dataset-level synthesis without calibration to a target population or urban system. \\
& P2 & Generation grounded in a target population through synthetic populations, surveys, census data, or population-level controls. \\
& P3 & Population-level execution with interaction, network loading, feedback, or other system dynamics. \\

\midrule
\textbf{Generation autonomy (A)}
& A0 & Generation requires extended observed histories of the target individual. \\
& A1 & Generation requires a partial trajectory, visit history, or other individual seed. \\
& A2 & Generation requires no observed individual history; samples are generated from learned priors and/or exogenous non-historical conditions, such as OD constraints, anchors, profiles, personas, or scenario specifications. \\
& A3 & Generation requires neither observed individual history nor manually predefined per-agent profiles; both synthetic agents or profiles and their mobility are produced from population and urban priors. \\
\bottomrule
\end{tabular}
\end{table}

\subsection{Behavioural meaning (M)}

Behavioural meaning (M) describes the extent to which synthetic human mobility represents interpretable behavioural structure or semantics rather than movement alone. This dimension concerns what the generated output says about why mobility occurs, what activity a trip serves, which mode is used, and how movement differs across people, households, or contexts. Behavioural meaning is recorded on two facets, because a single ladder would fuse two properties that the reviewed literature shows coming apart. The semantic side of this distinction is not new: modelling trajectories as semantically annotated rather than purely geometric objects is long established in the trajectory-analysis literature \citep{Parent2013}. What the facet adds here is the generation-side question of whether that semantics survives into the generated record or only guides its production. The first facet, behavioural semantics (M\textsubscript{b}), is whether behavioural meaning enters the generation process or is preserved in the generated record. The second, trajectory feasibility (M\textsubscript{f}), is whether generation is subject to explicit structural constraints. Neither concerns whether the output contains time information, since timestamps are part of most trajectory records. The two facets are scored separately and are not ordered against each other.

At M\textsubscript{b}0, the output is movement-only, such as coordinates, timestamps, grid cells, road links, or spatial tokens. Many trajectory-oriented generative models fall close to this level when their native output is a sequence of locations or coordinates \citep{Jiang2023,Long2023,ZhuDiffTraj2023}. At M\textsubscript{b}1, behavioural meaning is used as a condition or control signal. Activity purpose, transport mode, demographic group, user profile, persona, or contextual information may guide the generation process, but the final output remains a trajectory, token sequence, or mobility trace. In this case, meaning affects generation but is not preserved as an explicit field in the generated record. Conditional adversarial and profile-guided trajectory models often fall into this category \citep{Rao2023,Song2024,Li2025}. At M\textsubscript{b}2, behavioural meaning is explicit in the output itself. The generated record contains activity type, trip purpose, transport mode, persona, household role, or comparable behavioural fields. Activity-chain models, population-first pipelines, activity-based and agent-based simulations, and LLM-agentic travel diary generators are closer to this level when they natively produce activity plans, trip purposes, or semantically structured itineraries \citep{Pappalardo2018,Felbermair2020,Nayak2023,WangUrbanResidents2024,LiMobAgent2024,Liu2026}.

Trajectory feasibility is recorded independently of that scale. At M\textsubscript{f}0, generation is not subject to explicit network, routing, or physical feasibility constraints. At M\textsubscript{f}1, generation explicitly incorporates road topology, network loading, route choice, speed limits, or comparable feasibility checks; the generated result may still be primarily a movement trace, but it is one that could be executed on the network. Topology-constrained and controlled trajectory generators are typical examples \citep{Jiang2023,Zhu2024}. Origin--destination calibration is deliberately excluded from this facet: it constrains aggregate demand rather than the physical realisability of an individual record, and is captured by population grounding instead.

Separating the two facets makes their co-occurrence a question rather than a definition. Were structural constraint and behavioural semantics levels of one ladder, they would be mutually exclusive by construction, and no method could be recorded as achieving both. Whether meaning-explicit outputs can also be grounded into feasible trips and trajectories is treated in Section~\ref{sec:challenges} as an open challenge; recording the facets separately is what allows that question to be put to the classification rather than settled by the coding scheme.

This distinction matters because methods with similar architectures can support very different urban analytics tasks. A diffusion model that generates realistic GPS traces and an activity-based simulator that generates daily plans may both be ``generative'', but they represent different kinds of mobility knowledge. M\textsubscript{b}0 methods may be useful for trace imitation, augmentation, or route-level benchmarking, while M\textsubscript{b}1 methods provide more controllable trajectory generation. M\textsubscript{b}2 methods are more suitable when the goal is to interpret behaviour, compare population groups, test policy scenarios, or evaluate distributional effects. Behavioural meaning is therefore a necessary requirement for synthetic mobility as urban behaviour, but it is not sufficient on its own. Whether meaning-explicit outputs can be grounded into feasible trips and trajectories is treated later as an open challenge.

\subsection{Population grounding and scale (P)}

Population grounding and scale (P) describes whether synthetic mobility is linked to a target population and whether it can represent population-level mobility patterns. The operational question for P is whether the generated records are calibrated or grounded to a specified population, rather than merely produced as independent samples. This dimension is not simply about the number of generated samples. A model may generate many trajectories and still fail to represent a full population if those samples are not grounded in sociodemographic structure, household composition, or residential distribution. More comprehensive models may even consider dynamic population presence.

At P0, a method generates individual samples without explicit target-population grounding. Many trajectory or sequence generators fall near this level when they produce individual traces, trips, or token sequences without calibrating them to a city, region, or population structure. At P1, a method produces a synthetic dataset for release, augmentation, or benchmarking. The dataset may contain many records, but it does not necessarily represent a target population or urban system.

At P2, generation is grounded in population structure or population-level constraints. This may involve static synthetic populations, census or survey priors, demographic controls, household structure, aggregate flows, dynamic population distributions, or macro-level calibration. Population synthesis and survey-driven people-flow pipelines are typical examples, because they begin from synthetic persons, households, or population priors before generating activities, trips, or mobility records \citep{Beckman1996,GuoBhat2007,muller2010population,Felbermair2020,Kashiyama2024}. Aggregate and dynamic population studies also matter at this level because they define population-level patterns that synthetic mobility may need to reproduce, such as daytime population, ambient population, or time-varying population distribution \citep{Deville2014,Martin2015,Bergroth2022,Khodabandelou2019}.

At P3, synthetic mobility is embedded in system-level population simulation. Synthetic people, activities, trips, routes, or flows are executed in a simulated environment with interaction, feedback, network loading, congestion, or closed-loop dynamics. Activity-based and agent-based systems are closest to this level when they connect individual behaviour to system-level consequences \citep{Ronald2012,Zufle2023,Pang2020,Cai2021,Liu2026}.

The critical threshold for practical synthetic mobility is the move from P1 to P2. Below this threshold, a method may generate plausible records but not a plausible city. Above it, synthetic mobility begins to function as a representation of a target population and its movement across urban space and time. The remaining challenge is not only to generate population-grounded mobility, but also to make individual behaviour and aggregate population patterns consistent; this is treated later as an open challenge.

\subsection{Generation autonomy (A)}

Generation autonomy (A) describes how much information about a target individual is required at generation time. The operational question for A is how much person-specific information must be supplied at generation time, ranging from long observed histories to population-level priors only. This dimension concerns the inputs needed when new mobility is produced, not the data used during training or calibration. A model may be trained on real traces, surveys, or aggregate mobility statistics and still have high generation autonomy if it can generate mobility for new synthetic agents without requiring their observed personal histories. Conversely, a method may be highly generative in architecture but low in autonomy if it mainly extends, sanitises, or completes known user traces.

At A0, generation is user-history-dependent. The method relies on extended observed individual histories, repeated visits, long-term mobility traces, or user-specific records. This setting is common in personalised generation, trajectory transformation, and privacy-preserving publishing methods that preserve or sanitise mobility patterns for known individuals \citep{Wang2017,Zheng2022,Rao2023,Du2023}. These methods can be valuable for protecting or augmenting existing data, but they do not solve the problem of generating mobility for a new synthetic population.

At A1, generation depends on a partial history or seed. The input may be a partial trajectory, a known origin and destination, a short visit sequence, an incomplete daily record, or a constrained continuation context. This setting is common in prediction, completion, infilling, interpolation, or constrained trajectory generation. Transformer-based infilling models and constraint-aware trajectory generators often fall near this level when they require a partial sequence or seed condition to produce the output \citep{Hsu2024,Kobayashi2023}. A1 methods are useful when part of the mobility record is already known, but they remain less suitable for generating mobility for entirely new synthetic residents.

At A2, generation requires no observed trajectory history for the generated individual. The method may sample from learned priors alone, or may be conditioned on a synthetic profile, persona, demographic description, home/work anchor, or contextual constraint. This is common in profile-guided trajectory generation, persona-conditioned LLM workflows, and methods that begin with synthetic agents whose attributes are already specified \citep{Song2024,WangUrbanResidents2024,LiMobAgent2024,Liu2026}. A2 methods are more suitable for scenario-based synthesis than A0 or A1 methods, because the generated individuals need not correspond to observed users. Where conditioning is supplied, they still depend on the quality and realism of the supplied profiles, anchors, or personas.

At A3, generation is population-prior autonomous. The method generates both the synthetic agents and their mobility from population priors, surveys, census data, aggregate flows, urban structure, and scenario assumptions, without requiring observed histories or predefined profiles for each generated individual. Population-first synthesis pipelines are closest to this level when they synthesise the population structure itself before generating activities, trips, or mobility records \citep{Felbermair2020}. Survey-driven pipelines that draw their behavioural profiles from an existing travel survey remain at A2, because the profiles are supplied rather than synthesised \citep{Kashiyama2024}. A3 is the strongest form of generation autonomy because it removes the dependence on observed personal traces altogether, which is what makes new-city, future-population, or counterfactual scenario generation possible in principle. Whether the behavioural priors a method carries remain valid in a different city is a separate question, treated in Section~\ref{sec:governance}.

Generation autonomy should not be confused with model flexibility. An unconditional GAN or diffusion model that samples trajectories from noise may be highly autonomous in the narrow sense that it does not require a user's observed history, but it may still be low in behavioural meaning and population grounding. Conversely, an activity-based model may be highly autonomous and population-grounded, but still require substantial behavioural priors and calibration. The value of A therefore depends on how it combines with M and P. High autonomy is especially important for practical large-scale synthetic mobility, but it is not a guarantee of realism. The open challenge is how to replace observed individual histories with credible behavioural priors derived from synthetic populations, urban context, aggregate data, and validated persona or profile design.

\subsection{Capability profiles and method selection}
\label{sec:profiles}

The three requirements are most useful read together, as a profile across \textbf{M}, \textbf{P} and \textbf{A} rather than as separate scores. The groupings below describe a primary design orientation, not an exclusive category: a method may span more than one. Section~\ref{sec:families} gives each family's strengths and limitations and \ref{app:mpa} the levels assigned to every reviewed method; what follows is only the mechanism that places each grouping where it sits.

Aggregate methods---spatial interaction, trip distribution, crowd flow, dynamic population mapping---are population-grounded but semantically thin, because they do not generate individual routines, activity purposes, or trip-level decisions \citep{Wilson1970,McNally2000,OrtuzarWillumsen2011,Deville2014,Martin2015,Bergroth2022}. Trajectory-output generators invert this. They are strong on local spatiotemporal pattern but weak on population grounding, because they produce traces or datasets without necessarily representing a target urban population \citep{Jiang2023,ZhuDiffTraj2023,Long2023}; their autonomy depends on the generation setting, since unconditional generators need no history at generation time while completion-oriented models require a seed \citep{Hsu2024}. Population-first pipelines carry explicit behavioural meaning where activity type, trip purpose, mode, or household role survive into the record, and reach the highest autonomy only when the synthetic persons as well as their mobility are produced from population priors \citep{Felbermair2020,Kashiyama2024,Lu2026}. Agent-based systems extend those pipelines by executing the plans, and reach P3 only where interaction, network loading, or feedback are actually present rather than merely available \citep{Ronald2012,Zufle2023,Cai2021,Liu2026}. LLM-agentic workflows are semantically the richest and the most variable in grounding, which rises only where the workflow is connected to a synthetic population, aggregate calibration, or a simulation environment \citep{WangUrbanResidents2024,LiMobAgent2024,Zhang2024,Li2025,SalvadorLLMAgent2026,Liu2026}.

One asymmetry in these signatures is a finding rather than a limitation of the instrument. The framework separates the deep generative family least: more than two-thirds of its methods sit at exactly P1 with A2, a concentration not shared by the transformer-based generators, where the same combination accounts for under a third. Within that family it is behavioural semantics, not population grounding or autonomy, that distinguishes one method from another---the family has concentrated on generating realistic traces from a corpus, and comparatively little of it is calibrated to a target population.

Taken together, these profiles reveal an underdeveloped region in the current design space. Practical synthetic mobility requires meaning-explicit outputs, trajectory feasibility, population-grounded generation, and high generation autonomy simultaneously, while also remaining consistent with aggregate urban regularities. Existing methods often satisfy part of this requirement, but rarely all of it, and the combination of explicit behavioural semantics with trajectory feasibility is rarer still. In this corpus it occurs only where mobility is executed in a simulated environment rather than sampled from a learned model: the methods that combine the two route or load a network as part of generation, and each of them is also population-grounded at P3. This is a weaker result than it may appear, and worth stating plainly: the families that execute mobility on a network are the families that satisfy a feasibility criterion defined in those terms. The substantive observation is the converse one. Methods that generate semantically rich plans outside an execution environment---the population-first pipelines and most LLM-agentic workflows---ground those plans by post-processing rather than by constraining generation, and the deep generative family that dominates the recent literature does neither. That separation between semantic generation and feasible realisation is the coupling problem Section~\ref{sec:challenges} identifies, and it is visible here as a structural feature of the corpus rather than as an impression. Aggregate consistency is the one element of this requirement that the framework does not itself record; Section~\ref{sec:challenges} treats it as an open challenge, and Section~\ref{sec:limitations} explains why no common benchmark currently supports assessing it.

These profiles state the dimensions as properties of methods. Read in the other direction, they answer the question this review set out from: given an analytical need, what must a method deliver? Table~\ref{tab:selection} maps the needs enumerated in Section~\ref{sec:intro} onto minimum levels, and reports how many of the classified methods meet each combination.

\begin{table}[!ht]
\centering
\small
\caption{Minimum capability requirements for the analytical needs identified in Section~\ref{sec:intro}, with the number (\#) of classified methods meeting each combination. Counts are over the methods in \ref{app:mpa} and describe this corpus, not the field. Generation autonomy is deliberately absent: it is fixed by the data available locally rather than by the analytical need, and is discussed below.}
\label{tab:selection}
\begin{tabular}{p{0.225\linewidth} p{0.155\linewidth} c c c c p{0.255\linewidth}}
\toprule
\textbf{Analytical need} & \textbf{Output required} & \textbf{M\textsubscript{b}} & \textbf{M\textsubscript{f}} & \textbf{P} & \textbf{\#} & \textbf{Families} \\
\midrule
Exposure or route analysis & Individual trajectory & --- & 1 & $\geq$1 & 11 & all but population synthesis \\
\addlinespace[2pt]
Behavioural interpretation & Activity schedule & 2 & --- & $\geq$1 & 22 & all but transformer \\
\addlinespace[2pt]
Distributional analysis & Population-grounded mobility & 2 & --- & $\geq$2 & 12 & population synthesis, agent-based, mechanistic, LLM-agentic \\
\addlinespace[2pt]
Modal-shift and policy scenarios & Responsive system & 2 & 1 & 3 & 3 & agent-based, LLM-agentic \\
\addlinespace[2pt]
Infrastructure and demand planning & Aggregate flows & any & n/a & $\geq$2 & 10 & deep generative, mechanistic, transformer \\
\bottomrule
\end{tabular}
\end{table}

Three features of the table are worth stating. First, the requirements are not nested: exposure analysis needs feasibility and tolerates no behavioural meaning, while behavioural interpretation needs the reverse, so a method adequate for one may be inadequate for the other regardless of how sophisticated it is. Second, the counts fall sharply as requirements are combined, and the most demanding row---a scenario that must respond to a changed transport system---is met by the same three methods identified in the system-execution profile above. Third, generation autonomy does not appear, because it is not a property of the analytical need. It is fixed by what is available locally: an analyst holding individual histories for the population of interest may use any level, whereas an analyst in a city without trace data requires A2 or A3 whatever the task. Autonomy therefore acts as a filter applied after the table rather than a requirement read from it.

The table also makes the limits of the framework legible. It shortlists but does not rank: it carries no information on computational cost, calibration effort, input-data requirements, or the availability of an implementation, and none of the reviewed methods has been benchmarked against another under matched conditions (Section~\ref{sec:limitations}). What it supports is the first step of a selection process---narrowing six families and more than a hundred methods to a defensible handful---not the last. The next section turns from this capability gap to the open research challenges that make such integration difficult in practice.

\section{Open Challenges and Research Gaps}
\label{sec:challenges}

Previous sections show that synthetic human mobility generation has progressed along several methodological directions, but the field still faces a set of unresolved challenges that are not attributable to model accuracy alone. The central problem is cross-layer consistency. A method may produce plausible trajectories without explaining the activities behind them, or generate rich activity plans without grounding them into feasible movement, or reproduce aggregate mobility patterns while saying little about the individuals who produced them. These gaps matter because synthetic mobility is increasingly expected to support not only data release or trace imitation, but also urban analytics, policy simulation, and scenario-based planning. The following challenges summarise where current methods remain limited and where future work is most needed.

\subsection{Connecting semantic behaviour with feasible trajectories}
\label{sec:coupling}

A persistent challenge is to connect semantic mobility representations with physically feasible trajectories. Activity-based, population-first, and LLM-agentic approaches can generate meaningful daily plans, activity purposes, travel diaries, or persona-conditioned routines, but these outputs often require downstream grounding into locations, trips, routes, and travel times \citep{Kashiyama2024,WangUrbanResidents2024,LiMobAgent2024,Liu2026}. Trajectory-focused GAN, diffusion, and transformer models invert the problem: they can produce realistic-looking traces or mobility tokens, but their outputs often lack explicit activity purpose, trip meaning, or demographic interpretation \citep{Jiang2023,ZhuDiffTraj2023,Hsu2024}.

Closing this gap requires translation in both directions. Semantic plans must be resolved into feasible movement through place assignment, mode choice, routing, time-window enforcement, and network consistency checks; generated trajectories should in turn be interpretable in terms of activity purpose, travel context, and daily routine. Without this bidirectional link, synthetic mobility remains either semantically rich but spatially incomplete, or spatially detailed but behaviourally shallow. Act2Loc illustrates one staged solution, generating activity sequences first and then assigning locations through mechanistic models \citep{Liu2024Act2Loc}; such designs help preserve the distinction between behavioural organisation and spatial realisation, although consistency across the two representations still requires careful validation.

\subsection{Maintaining micro--macro consistency at population scale}

A second challenge is consistency between individual-level outputs and aggregate urban patterns. Many data-driven trajectory generators are evaluated on local spatiotemporal realism, such as trajectory shape, distance distribution, or temporal regularity. These metrics are important, but they do not guarantee that the synthetic population will reproduce correct OD demand, population distribution, inflow and outflow, link traffic counts, or temporal demand curves once many samples are aggregated \citep{Kapp2024,Kong2023}. Aggregate-first approaches face the mirror problem, and may reproduce OD matrices or flow patterns while providing little evidence that the underlying individual routines, trip chains, or trajectories are plausible.

This tension is especially important for urban analytics, where city-scale applications often depend on aggregate outcomes such as congestion, station usage, population exposure, service demand, or spatial inequality \citep{Basalamah2023,Pelletier2011,Chen2018,Moro2021}, while policy interpretation often requires individual-level behavioural structure. One route through the tension is to make the macro constraint part of the generator rather than a post-hoc check, as in diffusion models whose denoising is conditioned on dynamic population distribution \citep{LongDynPopDiff2026}. Synthesis methods therefore need calibration across scales: individual mobility should remain plausible under local inspection, while the population as a whole should also match known spatial, temporal, and network-level constraints. This remains difficult because fine-grained trajectory realism and macro-level calibration are often optimised separately.

\subsection{Scenario validity and policy-responsive generation}

Synthetic mobility is particularly valuable when it can be perturbed under hypothetical conditions. Urban researchers and planners often need to ask what would happen under land-use change, transport disruption, public-transit improvement, remote-work policy, demographic change, or disaster response. Several of these require generating mobility for people who were never observed. Scenario work therefore depends on behavioural priors standing in for individual histories, and those histories contain information that is difficult to replace: home--work anchors, recurring destinations, temporal habits, preferences, and constraints \citep{Felbermair2020,Kashiyama2024}. Personas and prompts extend the available priors \citep{WangUrbanResidents2024,LiMobAgent2024,Liu2026}, but profiles that make outputs more interpretable and controllable may also encode simplified or overly smooth assumptions about social groups and travel behaviour \citep{LutzPersona2025,HuPersona2024,YuSyntheticBias2023}. The requirement is not to avoid observed histories but to replace them with validated and diverse sources of behavioural heterogeneity. Current methods differ substantially in how they support such scenario reasoning. Activity-based and agent-based simulations can introduce interventions through behavioural rules, network conditions, or service changes, but often require substantial calibration and domain expertise \citep{Ronald2012,Pang2020,Cai2021}. LLM-based and agentic approaches offer high-level semantic control through prompts and personas, subject to the grounding problem of Section~\ref{sec:coupling}. Trajectory-output methods can sometimes condition on OD pairs, topology, profiles, or constraints, but their control is often local rather than system-level \citep{Song2024,Zhu2024,Hsu2024,Li2025}.

The open problem is scenario validity. A scenario should not only alter surface-level inputs; it should propagate consistently across daily activity schedules, trip and tour records, trajectories, and aggregate outputs. A transit disruption, for example, should affect mode choice, route feasibility, travel time, activity timing, and aggregate demand. A work-from-home scenario should alter activity schedules, trip frequency, spatial distribution, and temporal rhythms. Current methods rarely provide this kind of transparent and validated cross-representation scenario propagation.

\subsection{Transferability, access, and privacy governance}
\label{sec:governance}

Three concerns cut across the challenges above. None is a property of a single methodological family, and none is recorded by the capability dimensions of Section~\ref{sec:mpa}, but each bears on whether synthetic mobility becomes usable infrastructure rather than a methodological literature.

\paragraph{Transferability.}
Generation autonomy is defined by what a method does not require at generation time, not by where its behavioural priors remain valid. A generator calibrated on one city's population structure, activity patterns and travel costs has no guarantee of validity in a city with different density, transit provision, vehicle ownership, or household composition. Recent work makes this concrete rather than hypothetical: WorldMove builds per-location profiles from gridded population, points of interest and commuting flows and generates individual trajectories for more than 1,600 cities without local trace data \citep{yuan2026worldmove}, and cross-city mobility transformers transfer learned representations to cities unseen during training \citep{WangCOLA2024}. Both illustrate a trade-off rather than a solution. Transferability is bought with resolution: the priors that are globally available---gridded population, satellite-derived land use, POI density---support broad coverage precisely because they are coarse, and the individual realism attainable from them is correspondingly limited. Whether a foreign-calibrated generator is adequate is therefore an empirical question about a particular city and a particular analytical task, and no established protocol for answering it emerged from the reviewed literature.

\paragraph{Access.}
The constraints that motivate synthetic mobility are unevenly distributed. Commercial mobility datasets are often purchasable rather than unobtainable, which means the binding constraint for many groups is budget rather than availability. That asymmetry shapes what is studied and by whom: well-resourced institutions can license data that others cannot, and findings become difficult to reproduce outside the group that paid for the input. Synthetic mobility is unusual among responses to this problem because a generated dataset need not reproduce the records whose access was restricted. Whether it may then be released openly is a question of the licence under which the source data were obtained, and commercial agreements frequently extend to derivative works and model outputs; the possibility is therefore one that has to be negotiated rather than assumed. Where release is permitted, open synthetic resources lower a barrier that is economic rather than technical. This is a different argument from privacy protection: it concerns who is able to conduct urban mobility research at all, rather than whose records are exposed by it.

\paragraph{Privacy protection and its evasion.}
Synthetic generation reduces the direct exposure of individual records, and this is the property most often invoked when synthetic data is proposed for release. The protection is real but conditional: synthetic data does not automatically guarantee privacy, and generators fitted closely to individual records can reproduce identifying structure \citep{stadler2022synthetic}. A governance risk follows from the gap between the claim and the condition. Because ``synthetic'' reads as a categorical assurance, the label can be used to move a dataset past data-protection review rather than to satisfy it, and the assurance then travels further than the evidence supporting it. Section~\ref{sec:scope} treats privacy mechanisms as a design and evaluation concern rather than as an organising axis of this review, and that scope decision stands; the concern here is the consequence of releasing synthetic mobility, not the mechanism for protecting it. Reporting what a generator was trained on, what it was calibrated against, and what disclosure assessment was performed is a minimum condition for the assurance to carry meaning \citep{EuropeanDataProtectionSupervisor2025,Jordon2022}.

\subsection{Limitations of this review}
\label{sec:limitations}

The challenges above concern the methods reviewed. This review is itself bounded in three ways.

First, it does not benchmark or empirically validate the methods it compares. Evaluation in this field remains fragmented. Trajectory generators are often evaluated using geometric or statistical similarity metrics, while population-first and activity-based methods are often evaluated against survey marginals, OD patterns, or aggregate flows. LLM-agentic workflows may be assessed through plausibility, controllability, or qualitative inspection. These evaluation practices are not directly comparable, because different methods generate different layers and target different use cases, and current reviews already emphasise the difficulty of evaluating synthetic mobility utility and privacy jointly \citep{Kapp2024,Kong2023}. Shared datasets and benchmark tasks do exist: the Human Mobility Prediction Challenge releases large-scale mobility as \emph{grid-discretised} location sequences rather than raw trajectories, and several of the transformer models reviewed above were evaluated on it under matched conditions \citep{yabe2024enhancing,Kobayashi2023,SolatorioGeoFormer2023}. What does not exist is a benchmark spanning the representational layers this review compares. Evaluation on grid-level sequences cannot assess trajectory feasibility, which the discretisation removes by construction, and it says nothing about schedule semantics, population grounding, or aggregate fidelity. No common reporting standard exists against which the methods reviewed here could have been run under matched conditions across those layers. The comparisons in this paper, including the Meaning--Population--Autonomy assignments in Section~\ref{sec:mpa}, therefore rest on the design characteristics and native outputs that each method reports.

Second, the literature search draws on a single bibliographic database and covers predominantly English-language work, so coverage of preprint servers, of some conference proceedings, and of non-English research is incomplete. The query's generation terms are also trajectory- and mobility-synthesis phrases, so the population-synthesis, activity-based and travel-demand traditions were reached through citation-based expansion rather than through the database search itself. Beyond the database search and citation-based expansion described in Section~\ref{sec:litcollection}, a number of references were identified through the authors' domain knowledge of the field. The resulting corpus is best read as a representative sample of the design space rather than an exhaustive enumeration.

Third, the capability dimensions are coarse, and one of them is not a single ordered scale. Population grounding and scale combines two adjacent but distinct properties: P2 records whether generation is calibrated to a target population, whereas P3 records population-level execution with interaction, network loading, or feedback. A method can reach P3 without satisfying P2, as a microscopic traffic simulator does when it executes vehicles on a network with congestion feedback while taking its travel demand as exogenous \citep{Behrisch2011}. P3 is therefore better read as an adjacent capability than as a stronger form of P2, and a comparison of two methods across that boundary is not a comparison along one scale. The dimensions also discriminate unevenly between families. They separate the population-first and simulation traditions sharply, but leave most deep generative methods at the same coordinates, as Section~\ref{sec:profiles} notes, so the framework is a coarser instrument inside that family than across the field as a whole. Population grounding and generation autonomy are also less independent than presenting them as separate dimensions implies. Among the classified methods for which both are defined, almost half occupy a single cell, at P1 with A2; 23 of the 24 methods at A0 or A1 sit at P0 or P1, and seven of the eight at A3 sit at P2 or P3. The association is not an artefact of coding so much as a property of the field---a method that must be given the target individual's history is unlikely also to have been calibrated to a population it did not observe---but it means the two dimensions carry less independent information than their separate presentation suggests, and that readers should not treat a joint P and A profile as two independent pieces of evidence.

A stronger evaluation framework should align metrics with the intended urban-analytics task. At a minimum, synthetic mobility should be assessed across several dimensions: micro-level behavioural realism, trajectory and network feasibility, semantic consistency, macro-level fidelity, downstream utility, and privacy risk. Moving toward reporting standards that make methods comparable across representational layers and methodological families is a precondition for the kind of systematic comparison this review could not perform.

\section{Conclusion}
\label{sec:conclusion}
Synthetic human mobility generation has become increasingly important as urban researchers seek mobility data that can be shared, compared, and used for scenario-based analysis. Real mobility data remains indispensable, but it is often fragmented across surveys, passive digital traces, smart-card systems, and aggregate sensors, each with limitations in accessibility, coverage, semantic richness, and privacy risk. Synthetic human mobility data offers a possible response to these limitations, but only if generated data is useful for the urban questions it is meant to support.

This review reframes synthetic human mobility generation as a problem of matching analytical needs with appropriate output forms and methodological capabilities. Accordingly, it links core mobility output concepts and methodological families to the practical capabilities captured by the Meaning--Population--Autonomy framework. This use-oriented perspective allows methods to be compared not only by how they are implemented, but also by what they generate, what information they require, and which urban-analytics tasks they can support.

The review shows that no single approach is suitable for all synthetic mobility needs. Existing methods provide complementary but partial capabilities: some reproduce local movement patterns, some preserve behavioural structure, some are grounded in synthetic populations, and others operate primarily at the aggregate urban level. For many urban-analytics applications, however, these capabilities cannot be considered entirely in isolation. The relevant evidence may need to connect who moves, what activities generate travel, how trips are realised through space and time, and whether the resulting patterns remain plausible at population and urban-system scales.

The central gap is therefore not the absence of generative models, but the lack of approaches that can connect these capabilities in a transparent and evaluable way.

Addressing this gap will require a shift from isolated model development toward integrated and modular synthesis pipelines. Such pipelines should make explicit what is generated natively, what is used as conditioning information, what is derived through grounding or refinement, and how outputs are evaluated across behavioural, spatial, and aggregate levels. They should also connect synthetic population construction, daily activity scheduling, trip and tour generation, trajectory grounding, and aggregate validation more systematically than most current approaches do. This does not mean that a single model must solve every part of the problem. Rather, future work should clarify how different modules, data sources, and modelling traditions can be combined while preserving interpretability, feasibility, and population-level consistency.

Synthetic human mobility will be most valuable when it becomes more than artificial trajectory data. Its value for urban analytics lies in providing reusable, interpretable, scenario-ready representations of how people, activities, transport systems, and urban spaces interact. A use-oriented and capability-oriented view can help move the field toward synthetic mobility data that is not only realistic in appearance, but also meaningful, population-grounded, and useful for understanding cities in motion.

Three conditions will determine whether that shift produces usable infrastructure rather than a larger methodological literature. The first is reporting: comparison across representational layers requires methods to state what they generate natively, what they require at generation time, and what they were validated against, and no common standard for this yet exists. The second is transfer: priors calibrated in one city are not thereby valid in another, and the resolution cost of globally available priors means broad coverage and local realism currently trade against each other. The third is governance: synthetic mobility lowers an economic barrier to research when it is released openly, but ``synthetic'' is not by itself an assurance of privacy, and provenance must travel with a released dataset for that assurance to mean anything.

\section*{Declaration of generative AI and AI-assisted technologies in the manuscript preparation process}

During the preparation of this work the authors used ChatGPT and PicDoc in order to generate icon elements for conceptual figures and refine figure layout, and Claude in order to improve the language and readability of the manuscript text and to assist with the \LaTeX{} preparation of the document. After using these tools, the authors reviewed and edited the content as needed and take full responsibility for the content of the published article.

\appendix
\renewcommand{\thesection}{Appendix~\Alph{section}}

\renewcommand{\thetable}{\Alph{section}.\arabic{table}}
\renewcommand{\thefigure}{\Alph{section}.\arabic{figure}}
\makeatletter
\@addtoreset{table}{section}
\@addtoreset{figure}{section}
\makeatother

\section{Scopus search query}
\label{app:scopus}

The following Scopus query was used to search titles, abstracts, and keywords. It was executed on
31 July 2026 and returned 219 records.

\begin{verbatim}
TITLE-ABS-KEY (
  (
    ("human mobility" OR "urban mobility" OR
    (human OR people OR pedestrian* OR traveler* OR commuter*))
    AND
    ("trajectory generation" OR "trajectory synthesis" OR
    "synthetic trajectory" OR "synthetic mobility" OR
    "mobility synthesis" OR "activity generation" OR
    "synthetic human mobility")
    AND
    ("geospatial AI" OR GeoAI OR geospatial OR GIS OR
    "urban computing" OR "deep learning" OR transformer OR
    "large language model" OR LLM OR GAN OR VAE OR
    "diffusion model" OR "agent-based" OR ABM OR
    "agent model" OR "agent models" OR
    "differential privacy" OR "privacy-preserving" OR anonym*)
  )
)
AND PUBYEAR > 2007 AND PUBYEAR < 2027
\end{verbatim}

\section{Classification of reviewed generation methods in the MPA capability space}
\label{app:mpa}

Table~\ref{tab:mpa_classification} positions every generation method reviewed in Section~\ref{sec:families} within the Meaning--Population--Autonomy space, applying the operational criteria of Table~\ref{tab:mpa_dimensions} to each work's stated native output and stated conditioning requirements. Four conventions govern the table. First, a dash indicates that a dimension is not defined for that method rather than that it scores lowest. Behavioural semantics is recorded wherever a mobility record or flow is generated, and is not defined for a population without mobility. Trajectory feasibility and generation autonomy are properties of an individual mobility record, so neither applies to a population without mobility or to an aggregate flow or density surface. Second, generation autonomy is judged by what the method requires \emph{for the individual it generates}; a model trained on a corpus of observed trajectories is not thereby placed at A0, whereas a model that needs the target individual's own history is. Third, placements describe the configuration reported in the cited work, and several methods can be operated at a different level once coupled to a population synthesiser or a downstream grounding module. Fourth, a small number of works that are not themselves generators are listed because Section~\ref{sec:families} discusses them as methodological antecedents, as sources of aggregate output, or as population inputs; the native-output column states the reason in each case, and their levels describe what they produce rather than a generative capability.

Methods are grouped by the methodological families of Section~\ref{sec:families} and ordered chronologically within each family, so the table also reads as a timeline of how each family developed.

Works cited in this review that are not themselves generation methods---reviews, behavioural theory, empirical mobility analyses, data-processing methods, persona and policy documents, and general-purpose model architectures---are outside the scope of this table and are not listed.

\begin{longtable}{@{}p{0.215\linewidth} p{0.345\linewidth} >{\centering\arraybackslash}p{0.055\linewidth} >{\centering\arraybackslash}p{0.055\linewidth} >{\centering\arraybackslash}p{0.05\linewidth} >{\centering\arraybackslash}p{0.05\linewidth}@{}}
\caption{Classification of the generation methods reviewed in Section~\ref{sec:families}, with their predictive, aggregate and population antecedents, within the Meaning--Population--Autonomy capability space. A dash denotes a dimension that is not defined for the method, not a lowest score. Behavioural meaning is recorded on two facets, semantics (M\textsubscript{b}) and trajectory feasibility (M\textsubscript{f}).}
\label{tab:mpa_classification}\\
\toprule
\textbf{Method} & \textbf{Native output} & \textbf{M\textsubscript{b}} & \textbf{M\textsubscript{f}} & \textbf{P} & \textbf{A} \\
\midrule
\endfirsthead
\multicolumn{6}{@{}l}{\footnotesize\itshape Table~\ref{tab:mpa_classification} continued from previous page}\\
\toprule
\textbf{Method} & \textbf{Native output} & \textbf{M\textsubscript{b}} & \textbf{M\textsubscript{f}} & \textbf{P} & \textbf{A} \\
\midrule
\endhead
\midrule
\multicolumn{6}{r@{}}{\footnotesize\itshape continued on next page}\\
\endfoot
\bottomrule
\endlastfoot
\addlinespace
\multicolumn{6}{@{}l}{\textbf{Mechanistic and count-based models}}\\
\addlinespace[2pt]
\citet{Wilson1970} & Aggregate spatial-interaction flows & M\textsubscript{b}0 & --- & P2 & --- \\
\citet{McNally2000} & Zonal OD flows and assigned link volumes & M\textsubscript{b}0 & --- & P2 & --- \\
\citet{AshbrookStarner2002} & Predicted next location from clustered GPS significant places; listed as a predictive antecedent & M\textsubscript{b}0 & M\textsubscript{f}0 & P0 & A0 \\
\citet{OrtuzarWillumsen2011} & Zonal OD demand, mode shares and assigned flows & M\textsubscript{b}0 & --- & P2 & --- \\
\citet{Gambs2012} & Predicted next location; listed as a predictive antecedent & M\textsubscript{b}0 & M\textsubscript{f}0 & P0 & A0 \\
\citet{Mathew2012} & Predicted next location; listed as a predictive antecedent & M\textsubscript{b}0 & M\textsubscript{f}0 & P0 & A0 \\
\citet{Simini2012} & Aggregate commuting and migration flows & M\textsubscript{b}0 & --- & P2 & --- \\
\citet{Mir2013} & Synthetic call detail records for a synthetic population & M\textsubscript{b}1 & M\textsubscript{f}0 & P2 & A2 \\
\citet{Baratchi2014} & Stay-point and transition sequences & M\textsubscript{b}0 & M\textsubscript{f}0 & P0 & A0 \\
\citet{JiangTimeGeo2016} & Stay sequences with durations and daily mobility networks & M\textsubscript{b}1 & M\textsubscript{f}0 & P2 & A1 \\
\citet{Pang2017} & Reproduced daily travel behaviour sequences & M\textsubscript{b}1 & M\textsubscript{f}0 & P1 & A1 \\
\citet{Wang2017} & Perturbed personal trajectories & M\textsubscript{b}0 & M\textsubscript{f}0 & P1 & A0 \\
\citet{Gursoy2018} & Synthetic location traces & M\textsubscript{b}0 & M\textsubscript{f}0 & P1 & A2 \\
\citet{Pappalardo2018} & Mobility diary plus spatio-temporal trajectory & M\textsubscript{b}1 & M\textsubscript{f}0 & P1 & A2 \\
\citet{Qiao2018} & Predicted next location; listed as a predictive antecedent & M\textsubscript{b}0 & M\textsubscript{f}0 & P0 & A0 \\
\citet{Liu2019} & Synthetic trajectories from an n-gram model & M\textsubscript{b}0 & M\textsubscript{f}0 & P1 & A2 \\
\citet{Anda2021} & Daily schedules of stays with locations, times and durations & M\textsubscript{b}0 & M\textsubscript{f}0 & P2 & A3 \\
\citet{Cornacchia2021} & Individual trajectories (locations and times) & M\textsubscript{b}0 & M\textsubscript{f}0 & P1 & A2 \\
\citet{Yao2022} & Synthetic point trajectories under differential privacy & M\textsubscript{b}0 & M\textsubscript{f}0 & P1 & A2 \\
\citet{Du2023} & Synthetic point/grid trajectories & M\textsubscript{b}0 & M\textsubscript{f}0 & P1 & A2 \\
\citet{Hu2024} & Synthetic trajectory stream & M\textsubscript{b}0 & M\textsubscript{f}0 & P1 & A0 \\
\citet{Liu2024Act2Loc} & Activity sequences with assigned locations (synthetic trajectories) & M\textsubscript{b}2 & M\textsubscript{f}0 & P2 & A2 \\
\citet{WangOsaragi2024} & Daily activity sequences & M\textsubscript{b}2 & M\textsubscript{f}0 & P2 & A2 \\
\citet{Liu2025} & Individual daily activity schedules & M\textsubscript{b}2 & M\textsubscript{f}0 & P1 & A2 \\
\citet{LinGraphIRL2026} & Trajectories generated under incomplete observation & M\textsubscript{b}1 & M\textsubscript{f}1 & P1 & A2 \\
\addlinespace
\multicolumn{6}{@{}l}{\textbf{Population synthesis and survey-driven mobility pipelines}}\\
\addlinespace[2pt]
\citet{Beckman1996} & Synthetic baseline population & --- & --- & P2 & --- \\
\citet{Waddell2002} & Household and firm location choices, land-use and real-estate outcomes & --- & --- & P3 & --- \\
\citet{GuoBhat2007} & Synthetic population for travel microsimulation & --- & --- & P2 & --- \\
\citet{Sugiki2012} & Agent-based household micro-dataset & --- & --- & P2 & --- \\
\citet{BarthelemyToint2013} & Synthetic population (households and individuals) & --- & --- & P2 & --- \\
\citet{Ballis2020} & Home-based tours with trip purpose & M\textsubscript{b}2 & M\textsubscript{f}0 & P2 & A2 \\
\citet{Felbermair2020} & Synthetic population with activity chains and travel plans & M\textsubscript{b}2 & M\textsubscript{f}0 & P2 & A3 \\
\citet{MurataSyntheticPopulation} & Japanese synthetic population with protection levels & --- & --- & P2 & --- \\
\citet{Smolak2020} & Population-level mobility model for simulating shareable data & --- & --- & P2 & --- \\
\citet{Kajiwara2022} & Estimated and forecast household/population distribution; listed as a population input & --- & --- & P2 & --- \\
\citet{Wu2022GBSyntheticPopulation} & Synthetic individual-level micro-dataset for Great Britain & --- & --- & P2 & --- \\
\citet{UrbanPop2023} & Georeferenced synthetic agents at night-time and daytime locations & M\textsubscript{b}2 & M\textsubscript{f}0 & P2 & A3 \\
\citet{Kashiyama2024} & Nationwide synthetic daily mobility with activity labels & M\textsubscript{b}2 & M\textsubscript{f}0 & P2 & A2 \\
\citet{Li2025a} & Individualised pseudo-personal mobility records & M\textsubscript{b}1 & M\textsubscript{f}0 & P2 & A2 \\
\citet{Lu2026} & Individual spatiotemporal activity sequences & M\textsubscript{b}2 & M\textsubscript{f}0 & P2 & A3 \\
\addlinespace
\multicolumn{6}{@{}l}{\textbf{Activity-based and agent-based simulation}}\\
\addlinespace[2pt]
\citet{Behrisch2011} & Microscopic vehicle traffic simulation with routes and network loading & M\textsubscript{b}0 & M\textsubscript{f}1 & P3 & A2 \\
\citet{Ronald2012} & Scheduled social activities over a social network & M\textsubscript{b}2 & M\textsubscript{f}0 & P1 & A2 \\
\citet{Horni2016} & Daily activity programmes, trips, routes and link flows & M\textsubscript{b}2 & M\textsubscript{f}1 & P3 & A2 \\
\citet{Pang2020} & Simulated mass movement of individuals & M\textsubscript{b}1 & M\textsubscript{f}0 & P2 & A2 \\
\citet{Cai2021} & Simulated agent movement following home-work-home activity chains & M\textsubscript{b}2 & M\textsubscript{f}0 & P2 & A2 \\
\citet{Glake2022} & Learned spatio-temporal trajectories from a simulation system & M\textsubscript{b}1 & M\textsubscript{f}1 & P1 & A2 \\
\citet{Zufle2023} & Agent patterns of life: activities, places, social ties and wealth & M\textsubscript{b}2 & M\textsubscript{f}1 & P3 & A2 \\
\addlinespace
\multicolumn{6}{@{}l}{\textbf{Deep generative and trajectory-based models}}\\
\addlinespace[2pt]
\citet{LiuST-RNN2016} & Predicted next location; listed as a predictive antecedent & M\textsubscript{b}0 & M\textsubscript{f}0 & P0 & A0 \\
\citet{Kulkarni2017} & Synthetic trajectory dataset & M\textsubscript{b}0 & M\textsubscript{f}0 & P1 & A2 \\
\citet{ZhangSTResNet2017} & Citywide inflow/outflow per region; listed for its aggregate output layer & M\textsubscript{b}0 & --- & P2 & --- \\
\citet{kulkarni2018generative} & Synthetic mobility trajectory datasets & M\textsubscript{b}0 & M\textsubscript{f}0 & P1 & A2 \\
\citet{Feng2020} & Synthetic mobility location sequences & M\textsubscript{b}0 & M\textsubscript{f}0 & P1 & A2 \\
\citet{rao2020lstm} & Synthetic semantic trajectories with location category & M\textsubscript{b}2 & M\textsubscript{f}0 & P1 & A0 \\
\citet{Wang2021} & Fine-grained large-scale GPS trajectories & M\textsubscript{b}0 & M\textsubscript{f}1 & P1 & A2 \\
\citet{Berke2022} & Synthetic mobility traces for a synthetic population & M\textsubscript{b}0 & M\textsubscript{f}0 & P2 & A2 \\
\citet{YuanActivityTraj2022} & Activity trajectories with activity semantics & M\textsubscript{b}2 & M\textsubscript{f}0 & P1 & A2 \\
\citet{Zheng2022} & Synthetic trajectories resistant to social-relationship attacks & M\textsubscript{b}0 & M\textsubscript{f}0 & P1 & A0 \\
\citet{Gong2023} & Separate stopping-point and moving-point sequences & M\textsubscript{b}0 & M\textsubscript{f}0 & P1 & A2 \\
\citet{Huang2023} & Synthetic trajectories assembled from retrieved fragments & M\textsubscript{b}0 & M\textsubscript{f}0 & P1 & A2 \\
\citet{Jiang2023} & Continuous road-network trajectories & M\textsubscript{b}0 & M\textsubscript{f}1 & P1 & A2 \\
\citet{Long2023} & Synthetic human trajectories & M\textsubscript{b}0 & M\textsubscript{f}0 & P1 & A2 \\
\citet{Rao2023} & Synthetic trajectories from K-anonymised mobility matrices & M\textsubscript{b}1 & M\textsubscript{f}0 & P1 & A0 \\
\citet{Wang2023} & Synthetic mobility trajectories under privacy constraints & M\textsubscript{b}0 & M\textsubscript{f}0 & P1 & A2 \\
\citet{Zhang2023} & Modality-aware synthetic trajectories & M\textsubscript{b}1 & M\textsubscript{f}0 & P1 & A2 \\
\citet{ZhuDiffTraj2023} & Synthetic GPS trajectories & M\textsubscript{b}0 & M\textsubscript{f}0 & P1 & A2 \\
\citet{ZhuSynMob2023} & Released synthetic GPS trajectory dataset & M\textsubscript{b}0 & M\textsubscript{f}0 & P1 & A2 \\
\citet{chu2024simulating} & Synthetic trajectories from a diffusion model & M\textsubscript{b}0 & M\textsubscript{f}0 & P1 & A2 \\
\citet{Jia2024} & Synthetic trajectories & M\textsubscript{b}0 & M\textsubscript{f}0 & P1 & A2 \\
\citet{Netzler2024} & City-wide synthetic individual mobility from grid cells & M\textsubscript{b}0 & M\textsubscript{f}0 & P1 & A2 \\
\citet{Song2024} & Synthetic trajectories under a user profile & M\textsubscript{b}1 & M\textsubscript{f}0 & P1 & A2 \\
\citet{Tenzer2024} & Trajectories sampled from continuous latent fields & M\textsubscript{b}0 & M\textsubscript{f}0 & P1 & A2 \\
\citet{Zhu2024} & Road-constrained synthetic trajectories & M\textsubscript{b}0 & M\textsubscript{f}1 & P1 & A2 \\
\citet{Cao2025} & Neighbourhood-level synthetic trajectories & M\textsubscript{b}1 & M\textsubscript{f}0 & P1 & A2 \\
\citet{Crivellari2025} & Synthetic location sequences & M\textsubscript{b}0 & M\textsubscript{f}0 & P1 & A2 \\
\citet{Deng2025} & Trajectories generated under contextual conditions & M\textsubscript{b}1 & M\textsubscript{f}0 & P1 & A2 \\
\citet{Lakmal2025} & Controllable vehicle trajectories with semantic context & M\textsubscript{b}1 & M\textsubscript{f}0 & P1 & A2 \\
\citet{rong2026satellites} & Commuting origin-destination flow matrices & M\textsubscript{b}0 & --- & P2 & --- \\
\citet{xu2025predicting} & Fine-grained inter-regional mobility flows & M\textsubscript{b}0 & --- & P2 & --- \\
\citet{Yuan2025} & City-scale movements, population flows and their coupling & M\textsubscript{b}0 & M\textsubscript{f}0 & P2 & A2 \\
\citet{Zhang2025} & Synthetic urban mobility trajectories & M\textsubscript{b}1 & M\textsubscript{f}0 & P1 & A2 \\
\citet{BostanipourMobilityGraphs2026} & Synthetic individual mobility graphs with semantic location types & M\textsubscript{b}2 & M\textsubscript{f}0 & P1 & A2 \\
\citet{LongDynPopDiff2026} & Trajectories consistent with dynamic population distribution & M\textsubscript{b}0 & M\textsubscript{f}0 & P2 & A2 \\
\citet{Luo2026} & Synthetic trajectories & M\textsubscript{b}1 & M\textsubscript{f}0 & P1 & A2 \\
\citet{wang2026sat2flow} & Structurally coherent OD flow matrices & M\textsubscript{b}0 & --- & P2 & --- \\
\citet{xu2026multimodal} & Large-scale human mobility flow matrices & M\textsubscript{b}0 & --- & P2 & --- \\
\citet{XuGeoGen2026} & Fine-grained LBSN check-ins with POI and timestamp & M\textsubscript{b}1 & M\textsubscript{f}0 & P1 & A2 \\
\citet{yuan2026worldmove} & City-scale individual movement trajectories (1,600+ cities) & M\textsubscript{b}1 & M\textsubscript{f}0 & P2 & A3 \\
\citet{Zou2026} & Synthetic mobile-signaling trajectories & M\textsubscript{b}0 & M\textsubscript{f}0 & P1 & A2 \\
\addlinespace
\multicolumn{6}{@{}l}{\textbf{Transformer-based mobility language models}}\\
\addlinespace[2pt]
\citet{FengDeepMove2018} & Predicted next location; listed as a predictive antecedent & M\textsubscript{b}0 & M\textsubscript{f}0 & P0 & A0 \\
\citet{MizunoGPT2Traj2022} & Generated daily trajectories as grid-code token sequences & M\textsubscript{b}0 & M\textsubscript{f}0 & P1 & A2 \\
\citet{XueLangFM2022} & Forecast POI visitor flows; listed for its aggregate output layer & M\textsubscript{b}0 & --- & P1 & --- \\
\citet{CorriasTransformerGCN2023} & Predicted next place; listed as a predictive antecedent & M\textsubscript{b}0 & M\textsubscript{f}0 & P0 & A0 \\
\citet{Kobayashi2023} & Generated grid-cell location sequences & M\textsubscript{b}1 & M\textsubscript{f}0 & P1 & A1 \\
\citet{SolatorioGeoFormer2023} & Forecast continuation of a mobility sequence; listed as a predictive antecedent & M\textsubscript{b}0 & M\textsubscript{f}0 & P1 & A1 \\
\citet{Hsu2024} & Completed sequences of visits with times & M\textsubscript{b}0 & M\textsubscript{f}0 & P1 & A1 \\
\citet{LiUrbanGPT2024} & Spatio-temporal forecasts across urban tasks; listed for its aggregate output layer & M\textsubscript{b}0 & --- & P2 & --- \\
\citet{WangCOLA2024} & Simulated human trajectories, transferable across cities & M\textsubscript{b}0 & M\textsubscript{f}0 & P1 & A2 \\
\citet{WangOsaragiTransformer2024} & Synthesised daily movement sequences & M\textsubscript{b}0 & M\textsubscript{f}0 & P1 & A2 \\
\citet{Yasuda2025} & Predicted continuation of a mobility trajectory; listed as a predictive antecedent & M\textsubscript{b}0 & M\textsubscript{f}0 & P1 & A1 \\
\citet{Haydari20261681} & Synthetic mobility token sequences on a road network & M\textsubscript{b}0 & M\textsubscript{f}1 & P1 & A2 \\
\citet{WuPretrainedMobility2026} & Pretrained mobility representations and generated location-token sequences & M\textsubscript{b}0 & M\textsubscript{f}0 & P2 & A2 \\
\addlinespace
\multicolumn{6}{@{}l}{\textbf{Large language models and agentic workflows}}\\
\addlinespace[2pt]
\citet{Park2023} & Agent daily routines: activities, plans, conversations and movements & M\textsubscript{b}2 & M\textsubscript{f}0 & P0 & A2 \\
\citet{WangLLMMob2023} & Predicted next location; listed as a predictive antecedent & M\textsubscript{b}0 & M\textsubscript{f}0 & P0 & A0 \\
\citet{LiMobAgent2024} & Travel diaries with activities and reasons & M\textsubscript{b}2 & M\textsubscript{f}0 & P1 & A2 \\
\citet{WangUrbanResidents2024} & Personal activity sequences with motivations & M\textsubscript{b}2 & M\textsubscript{f}0 & P1 & A2 \\
\citet{Zhang2024} & Mobility records with activity purpose and transport mode & M\textsubscript{b}2 & M\textsubscript{f}0 & P1 & A2 \\
\citet{ChenInteractiveNextLoc2025} & Predicted next location, with dialogue-based rationale; listed as a predictive antecedent & M\textsubscript{b}0 & M\textsubscript{f}0 & P0 & A0 \\
\citet{JuTrajLLM2025} & Simulated trajectories with generated personas and selected activities & M\textsubscript{b}2 & M\textsubscript{f}0 & P1 & A3 \\
\citet{Li2025} & Trajectories satisfying specified visit constraints & M\textsubscript{b}1 & M\textsubscript{f}0 & P1 & A2 \\
\citet{Li2025b} & Synthetic mobility trajectories from a hierarchical LLM agent workflow & M\textsubscript{b}1 & M\textsubscript{f}0 & P1 & A2 \\
\citet{WangAgenticRoute2025} & Simulated day-to-day route choices & M\textsubscript{b}1 & M\textsubscript{f}1 & P1 & A2 \\
\citet{Liu2026} & Agent activity plans and adaptive travel behaviour in a simulation environment & M\textsubscript{b}2 & M\textsubscript{f}1 & P3 & A3 \\
\citet{SalvadorLLMAgent2026} & Synthetic household travel-survey responses & M\textsubscript{b}2 & M\textsubscript{f}0 & P2 & A3 \\
\addlinespace
\multicolumn{6}{@{}l}{\textbf{Dynamic population distribution (calibration and validation targets)}}\\
\addlinespace[2pt]
\citet{Dobson2000LandScan} & Global ambient population density grid & --- & --- & P2 & --- \\
\citet{Terada2013MSS} & Estimated population distribution from mobile network data & --- & --- & P2 & --- \\
\citet{Deville2014} & Dynamic population density maps & --- & --- & P2 & --- \\
\citet{Martin2015} & Time-specific gridded population estimates & --- & --- & P2 & --- \\
\citet{Tatem2017WorldPop} & Open high-resolution gridded population and demographics & --- & --- & P2 & --- \\
\citet{Khodabandelou2019} & Static and time-varying urban population density & --- & --- & P2 & --- \\
\citet{Population247NRT2021} & Time-specific gridded population by age group, England & --- & --- & P2 & --- \\
\citet{Bergroth2022} & Hourly 250 m population distribution grid, Helsinki & --- & --- & P2 & --- \\
\end{longtable}


\bibliography{sample}

\end{document}